\documentclass[USenglish,twocolumn]{article}

\usepackage[utf8]{inputenc}				
\usepackage[big,online]{dgruyter}	
\usepackage{lmodern} 
\usepackage{microtype}
\usepackage[numbers,square,sort&compress]{natbib}
\usepackage{blindtext}
\usepackage{hyperref}
\usepackage{cleveref}
\crefname{figure}{Figure}{Figures} 
\usepackage{graphicx}
\usepackage{caption}
\usepackage{subcaption}
\usepackage{comment}
\usepackage[hypcap=false]{caption}

\theoremstyle{dgthm}

\theoremstyle{dgdef}

\begin{document}

	\articletype{Research Article}
	\received{Month	DD, YYYY}
	\revised{Month	DD, YYYY}
  \accepted{Month	DD, YYYY}
  \journalname{De Gruyter Journal}
  \journalyear{YYYY}
  \journalvolume{XX}
  \journalissue{X}
  \startpage{1}
  \aop
  \DOI{10.1515/sample-YYYY-XXXX}

\title{General model for estimating range variances of terrestrial laser scanners based on (un-)scaled intensity values}

\runningtitle{In-situ intensity-based range variance model estimation}

\author*[1]{Omar AbdelGafar}
\author[2]{Selin Palaz}
\author[2]{Yihui Yang}
\author[2]{Christoph Holst} 
\runningauthor{O.AbdelGafar and C. Holst}
\affil[1]{\protect\raggedright 
Chair of Engineering Geodesy, TUM School of Engineering and Design, Technical University of Munich, 80333 Munich, Germany, E-mail: o.abdelgafar@tum.de}
\affil[2]{\protect\raggedright 
Chair of Engineering Geodesy, TUM School of Engineering and Design, Technical University of Munich, 80333 Munich, Germany}
	
\abstract{Recent advancements in technology have established terrestrial laser scanners (TLS) as a powerful instrument in geodetic deformation analysis. As TLS becomes increasingly integrated into this field, it is essential to develop a comprehensive stochastic model that accurately captures the measurement uncertainties. A key component of this model is the construction of a complete and valid variance-covariance matrix (VCM) for TLS polar measurements, which requires the estimation of variances for range, vertical, and horizontal angles, as well as their correlations. While angular variances can be obtained from manufacturer specifications, the range variance varies with different intensity measurements. As a primary contribution, this study presents an effective methodology for measuring and estimating TLS range variances using both raw and scaled intensity values. A two-dimensional scanning approach is applied to both controlled targets and arbitrary objects using TLS instruments that provide raw intensity values (e.g., Z+F~Imager~5016A) and those that output scaled intensities (e.g.,  Leica~ScanStation~P50). The methodology is further evaluated using field observations on a water dam surface. Overall, this work introduces a comprehensive workflow for modeling range uncertainties in high-end TLS systems.}

\keywords{
TLS, Stochastic Model, Variance Propagation, Calibration, Uncertainty, Scaling Function, Random errors, Point Clouds}

\maketitle
\section{Introduction} \label{introduction}
In present geodetic research and applications, terrestrial laser scanners (TLS) represent an important technological advancement, facilitating area-wise acquisition of 3D point clouds with both high precision and spatial resolution. Due to recent developments in laser technology and data processing algorithms, their capabilities have been significantly elevated, enabling the generation of dense point clouds on complex surfaces \citep{li_comparison_2021} and built structures \citep{zhou_high-precision_2024}.
These advancements strengthen the role of TLS in geodetic engineering applications, which require millimeter-level accuracy \citep{Holst_tlsdefo_2016}, particularly for deformation monitoring and structural analysis of critical infrastructures such as water dams \citep{kerekes_determining_2021}, bridges \citep{zhou_high-precision_2024}, and radio telescopes \cite{holst_terrestrial_2017,1779046}. Therefore, differentiating between the deformation components and the measurement uncertainties is vital \citep{yang2025find}.

The measurement product equals a discrete point cloud defined by $n$ local Cartesian coordinates and backscattered intensities $I$. However, these coordinates are originally expressed in polar form, consisting of ranges $r_i$ from the scanner to the scanned object, vertical angles $\theta_i$, and horizontal angles $\varphi_i$ of number $i=1,...n$. These measurements are affected by four -- partially highly-correlated -- error sources: scanner imperfections, atmospheric conditions, scanning geometry, and physical properties of the scanned surface \citep{soudarissanane_scanning_2011}. Thus, understanding the stochastic behaviors of TLS implies building up a fully propagated variance-covariance matrix (VCM) for the raw observations. 

To construct the VCM, it is necessary to represent the variances and covariances of these polar measurements, as illustrated in \autoref{fig:VCM}. The majority of the studies focus on the estimation of range variances \cite{wujanz_intensity-based_2017,wujanz_bestimmung_2018,heinz_strategy_2018} under controlled laboratory environments using specialized specimens and specific experimental configurations such as rail-bound comparator tracks, which inherently constrain the applicability of these methods. Moreover, these studies considered the relationship between raw intensity values and range uncertainties; however, raw intensity values are not provided by all TLS manufacturers, as the majority provide only scaled intensity values, which are generated through a scaling function applied during the data export process. The internal parameters and implementation of this function remain inaccessible, and the scaling cannot be switched off. While this can be useful for enhancing visualization or segmentation processes \citep{schmitz_how_2019,kaasalainen_radiometric_2009}, it affects direct range variance estimation. This issue was addressed in \cite{schmitz_how_2019}; nevertheless, the proposed method again relies on a specialized specimen.

\begin{center}
\includegraphics[width=\linewidth]{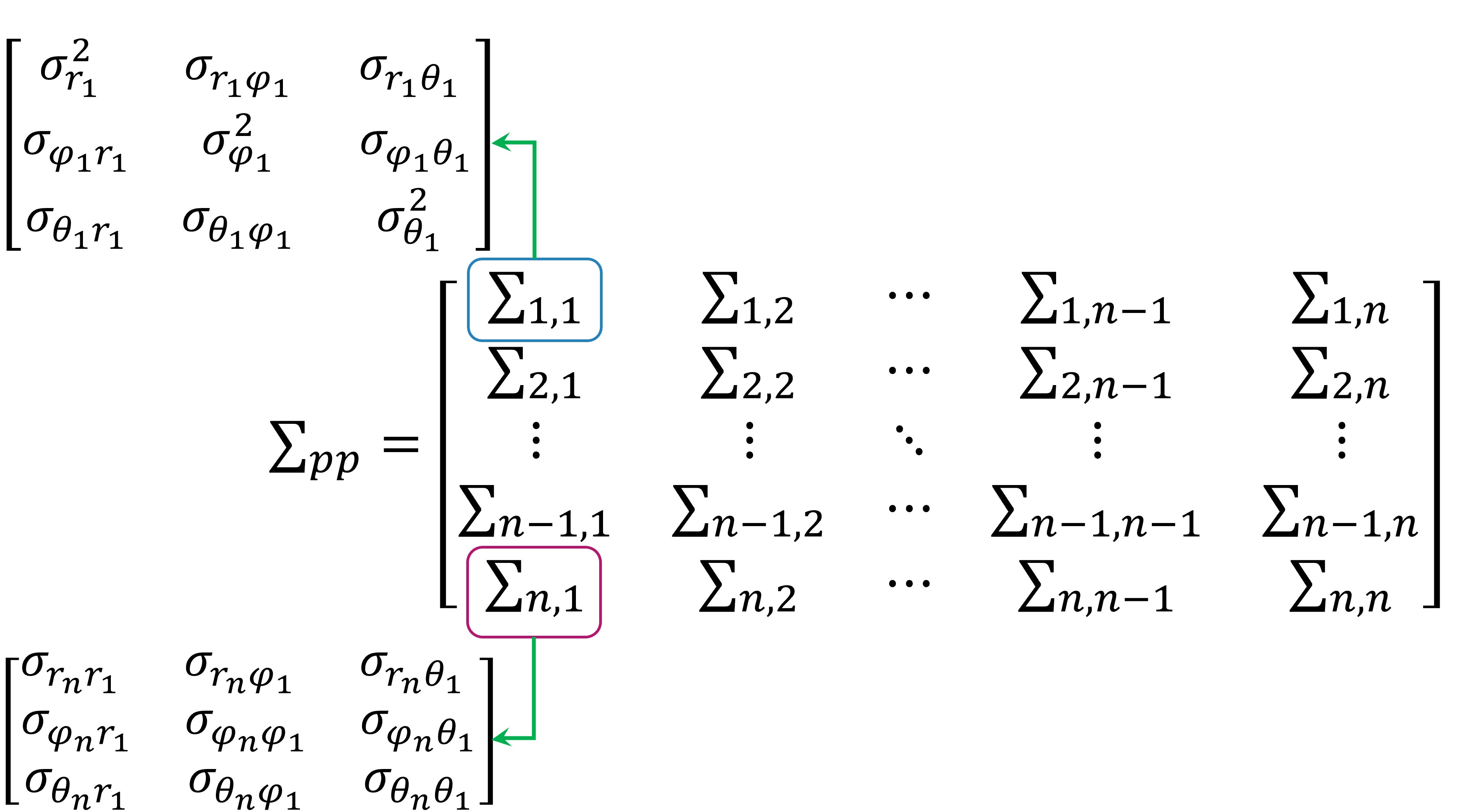}
\captionof{figure}{A fully propagated VCM of the polar observations of a terrestrial laser scan with $n$ points. $\Sigma_{1,1}$ (blue box) represents the VCM for one single scan point with number $1$, while $\Sigma_{n,1}$ (red box) represents the relative VCM between scan points $n$ and $1$.}
\label{fig:VCM}
\end{center}
To overcome these limitations, we address the following key contributions within this study: 
\begin{itemize}
\item We propose and evaluate a workflow for the in-situ estimation of range variances of high-end 3D laser scanners, utilizing a 2D scanning mode on arbitrary surfaces.
\item We investigate the impact of using both raw (unscaled) and scaled intensity values on the estimation of the intensity-based range variance model.
\item We introduce a general intensity-based range variance model to overcome the limitations imposed by intensity scaling.
\end{itemize}

The rest of this article is organized as follows: Section \ref{State of the art} provides an overview of previous studies on range uncertainty estimation. Section \ref{Methodology} describes the implemented methodological workflow. Section \ref{Experimental Setup} explains the experimental setup and data collection for evaluation. Section \ref{Results and Model Validation} presents the results obtained from applying the workflow to both raw and scaled intensity data sets and discusses their evaluation, followed by conclusions in Section \ref{conclusion}.

\section{Related works} \label{State of the art}

We split the related works into strategies for estimating range variances $\sigma^2_r$ based on raw intensity values (Section 2.1) and scaled intensity values (Section 2.2) and a summary (Section 2.3). Due to the study focus on range variances $\sigma^2_r$ , we skip a detailed review of methods for estimating other parameters of the VCM represented in \autoref{fig:VCM}. For further insights thereupon, see \citep{JurekKuhlmannHolst+2017+143+155,kerekes2023elementary}.

\subsection{Range variance estimation using raw intensity values}

Wujanz et al. (2017) \citep{wujanz_intensity-based_2017} proposed a strategy for empirically estimating range variances, $\sigma^2_r$, of laser scanner measurements in relationship to the backscattered intensity $I$ through an exponential function:
\begin{equation}
    \sigma_r = aI^b + c. 
    \label{eq:Intensity_based_range_variance_model}
\end{equation}
Here,  \(a\), \(b\), and \(c\) represent the parameters to be estimated.

The measured intensity values mainly depend on the distance between the object and the scanner, the incidence angle of the laser beam, and the properties of the scanned surface. 1D scanning mode was applied to estimate the intensity-based range variance parameters $a,b,c$, where repetitive single-point measurements were acquired at different distances and with different target materials \citep{wujanz_intensity-based_2017}. However, the use of the 1D mode is restricted due to safety concerns associated with the concentrated energy of the laser \citep{heinz_strategy_2018}. 

To overcome the limitation of the 1D mode, further studies investigated range noise by applying the standard 3D scanning mode \citep{wujanz_bestimmung_2018,hobiger_empirical_2018}. Their methods estimate residuals from best-fitting planes derived from the point clouds of planar targets that are oriented to ensure zero incidence angles and minimize angular encoder deviations. Nevertheless, these approaches rely on specific model assumptions that are not compatible with non-uniform backscattering, and thus are only applicable for specific experimental setups.

Based on the mentioned methodologies, Heinz et al. (2018) \citep{heinz_strategy_2018} introduced an approach, which is suitable for 2D laser scanners, i.e., profile laser scanners. Following this strategy, Schill et al. (2024) \citep{schill_intensity-based_2024} applied the method using high-end 3D laser scanners. The approach involves repetitive profile scanning of different targets with various scanning orientations to estimate range uncertainties without any geometric primitive assumptions.

\subsection{Range variance estimation using scaled intensity values}
\label{Scaled Intensity Values in Range Variance Estimation}

Most studies mentioned above only took raw intensity values to estimate the intensity-based range variance. However, not all laser scanner manufacturers provide these unscaled intensity data; most only offer scaled intensity values. The reason behind scaling intensity originates from airborne laser scanning applications, where the backscattered values need to be calibrated to improve the intensity-based data processing, like feature extraction and point cloud segmentation \citep{HOFLE2007415}. 

Many approaches have been proposed to calibrate and scale intensity values based on scanning geometry, such as range and incidence angles \citep{luzum2004normalizing, Yan02072016, Charaniya2004supervised} according to the Laser Radar Equation \cite{jelalian_laser_1992}:
\begin{equation}
   I_r = \frac{I_t \, D_r^2 \, \rho}{4 r^2} \, n_{\mathrm{sys}} \, n_{\mathrm{atm}} \, \cos(\theta)
\label{eq:radar_range_equation}
\end{equation} 
\noindent
where:
\begin{itemize}
    \item[$I_r$] received intensity at the detector, 
    \item[$I_t$] transmitted laser intensity,
    \item[$D_r$] effective receiver aperture diameter,
    \item[$\rho$] target reflectivity (backscatter coefficient),
    \item[$r$] range (distance between the scanner and the target),
    \item[$n_{\mathrm{sys}}$] system transmission efficiency (optical and electronic losses),
    \item[$n_{\mathrm{atm}}$] atmospheric transmission factor (two-way path attenuation),
    \item[$\theta$] incidence angle between the laser beam and the surface normal.
\end{itemize}
However, individual scaling functions applied by manufacturers remain proprietary and are not publicly available. Despite its benefits for point cloud classification and segmentation, the estimation of range variance based on these scaled intensities becomes more challenging.

Schmitz et al. (2019) \citep{schmitz_how_2019} first investigated the effect of scaled intensity values on range uncertainty. In this study, planar targets were scanned using the 3D mode, and residuals were then estimated from the fitted planes. This approach still introduces specific model assumptions, similar to \citep
{hobiger_empirical_2018, wujanz_bestimmung_2018}. On the other hand, the results highlight the impact of the scaling function on range uncertainty estimation, demonstrating that the intensity-based range variance model varies for different measuring distances.

\subsection{Conclusion of related works}
Although many strategies exist for determining intensity-based range variances for TLS, these methods (i) are mostly limited to scaled intensities and (ii) experience complications from the complexity of the measurement setup and the necessity of special specimens or controlled environments. To tackle these limitations (i) and (ii), a simple and efficient workflow is introduced in this article, enabling scanner users to estimate range uncertainties under casual conditions without the need for specific targets. Furthermore, both raw (unscaled) and scaled intensity values can be incorporated in this intensity-based range variance model, allowing the method to be applied to a larger variety of laser scanners.

\section{Methodology}
\label{Methodology}

We adopt the 2D scanning mode, applied to data sets that include either raw or scaled intensity values. The workflow is divided into two main steps: data pre-processing (Section \ref{Workflow for data preprocessing}) and model fitting (Section \ref{Workflow for model fitting}). Since the intensity scaling function affects the behavior of the estimated model compared to that obtained using raw intensity values, Section \ref{General model for scaled intensity} proposes a method to overcome this effect.

Our methodology does not depend on a specific measurement procedure, apart from scanning 2D profiles comprising at least 3000 profile lines. To demonstrate the general applicability of the methodology, Section \ref{Results and Model Validation} evaluates its performance based on both laboratory and field measurements, as outlined in Section \ref{Experimental Setup}.
\subsection{Data pre-processing}
\label{Workflow for data preprocessing}

In this step, the goal is to prepare the TLS data for the subsequent model fitting procedure. As illustrated in \autoref{fig:Methodology_workflow}, the process starts with exporting the polar observations together with the intensity values. This export is always performed using the manufacturer's software. Depending on the software, intensity values can be exported as raw values, as in the case of \textit{Z+F LaserControl} \citep{zf_lasercontrol}, or they may be subjected to a scaling function, resulting in scaled intensity values, as done by the \textit{Leica Profiling Software API} \citep{LeicaSoftware2025}. For other manufacturers, respective software packages should be used.

Following data export, outliers in both range and intensity values are identified and removed. To achieve this, both range and intensity values are grouped by the vertical ticks across all profiles of the measured target. Therefore, we ensure having a series of range values along with their corresponding intensity values.

\twocolumn[{
\begin{center}
\includegraphics[width=0.87\textwidth]{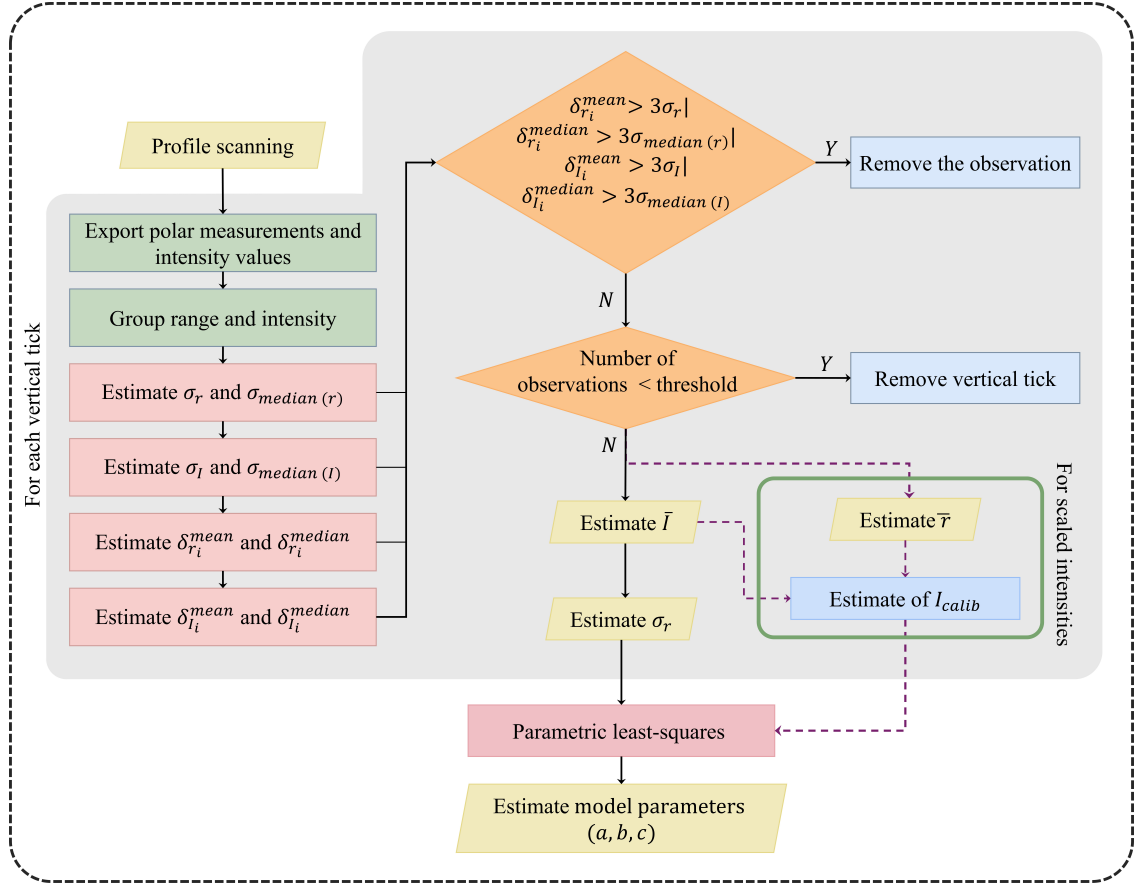}
\captionof{figure}{Schematic illustration of the complete workflow for estimating the intensity-based range variance model. The diagram illustrates the pre-processing of 2D profile-scanning data sets and the subsequent parameter estimation of the model using parametric least squares adjustment. The green box highlights the inputs used for estimating a generalized model based solely on scaled intensity values.}
\label{fig:Methodology_workflow}
\end{center}
\vspace{1em}
}]

Once the grouping is applied, the variability within each vertical tick is assessed by estimating the standard deviation $\sigma_{y}$ with respect to the mean $\bar{y}$, as well as the standard deviation $\sigma_{\text{median}(y)}$ with respect to the median $\text{median}(y)$, both for range $r_i$ as well as intensity values $I_i$ with $i$ being the number of points per tick:
\begin{equation}
    \sigma_{y} = \sqrt{\frac{1}{n-1} \sum_{i=1}^{n} (y_i - \bar{y})^2},
    \label{eq:std relative to mean}
\end{equation}
\begin{equation}
    \bar{y} = \frac{1}{n} \sum_{i=1}^{n} y_i,
    \label{eq:arithmetic mean}
\end{equation}
\begin{equation}
    \sigma_{\text{median}(y)} = \sqrt{\frac
{1}{n-1} \sum_{i=1}^{n} (y_i - \text{median}(y))^2}.
    \label{eq:std_median}
\end{equation}

Introducing a confidence level of 99.7\%, an observation will be classified as an outlier if its deviation from either the mean or the median is greater than three times the standard deviation of the mean or the median. This condition is expressed as:

 \begin{align}
    \delta_{y_i}^{\mathrm{mean}} &> 3 \sigma_{y} , \\
    \delta_{y_i}^{\mathrm{median}} \ &> 3 \sigma_{\text{median}(y)}.
    \label{eq:outlier_criteria}
\end{align}

Here, \( \delta_{y_i}^{\mathrm{mean}} \) and \( \delta_{y_i}^{\mathrm{median}} \) denote the absolute deviations from the mean and the median, respectively, and are defined as: 

 \begin{align}
    \delta_{y_i}^{\mathrm{mean}} &= | y_i - \bar{y} |, \\
    \delta_{y_i}^{\mathrm{median}} &= | y_i - \text{median}(y) |.
    \label{eq:deviation from the mean and the median}
\end{align}

Both the mean and the median are used to handle different outlier characteristics within each vertical tick. The median provides robustness against impulsive deviations, while the mean represents the overall statistical trend of the data, ensuring the removal of extreme outliers without distorting the local distribution. Therefore, all observations satisfying this condition are excluded from the vertical tick group, ensuring that only values within an acceptable range will be analyzed.

Additionally, to ensure that all vertical ticks have a sufficient number of observations, another constraint is applied: The vertical tick will be discarded if its number of observations is lower than a certain threshold defined based on the target scan data set. 
After those pre-checks eliminating individual observations or complete vertical ticks, the approved observations are taken to estimate averaged intensity values $\bar{I}$ using Eq. (\ref{eq:arithmetic mean}) as well as standard deviations for the ranges $\sigma_{r}$ using Eq. (\ref{eq:std relative to mean}), all per vertical tick.

\subsection{Intensity calibration for scaled intensity values}
\label{General model for scaled intensity}

By applying this pre-processing workflow to data sets with raw intensity values, it can be observed that, for each scanning rate, the data acquired at different distances are aligned and follow a consistent trend (\autoref{fig: Z+F_3_scanning rates data sets}). This outcome is in agreement with previous studies \citep{schill_intensity-based_2024}. Contrary, for scaled intensity values, the points only follow the same trend within individual scanning rates. For varying scanning rates, this trend varies with distance (\autoref{fig: ZF+Imager_10m_25m_50m_preprocessed_data sets_scaled_Intensity}).

\begin{center}
\includegraphics[width=1\linewidth]{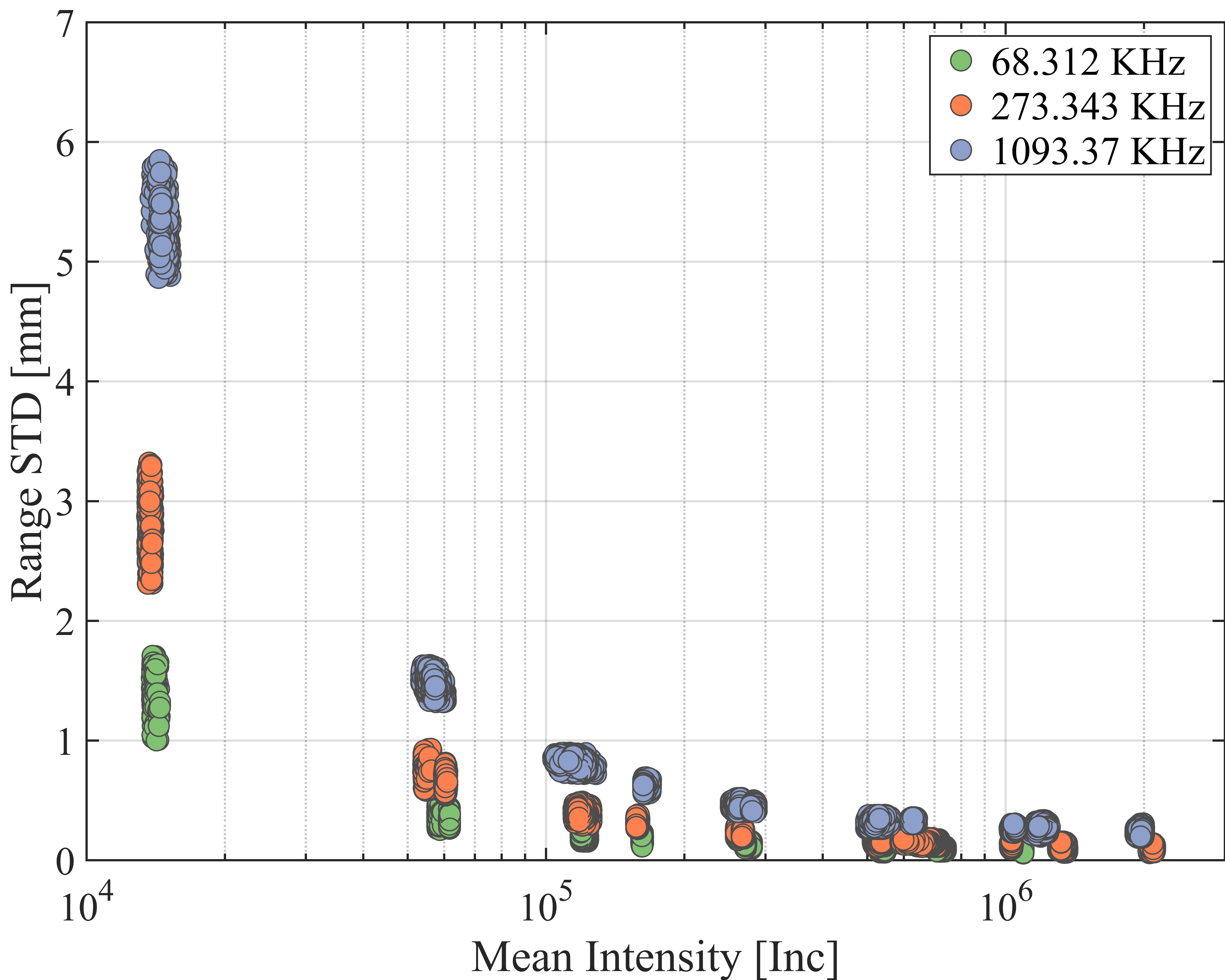}
\captionof{figure}{Exemplary relationship for Z+F~Imager~5016A between range uncertainty and mean raw intensity for the Spectralon board observed at distances of 10~m, 25~m, and 50~m for three different scanning rates: 68.312~kHz (green), 273.343~kHz (orange), and 1093.37~kHz (blue). Each scanning rate (in the same color) contains different measuring distances.}
\label{fig: Z+F_3_scanning rates data sets}
\end{center}

\begin{center}
\includegraphics[width=\linewidth]{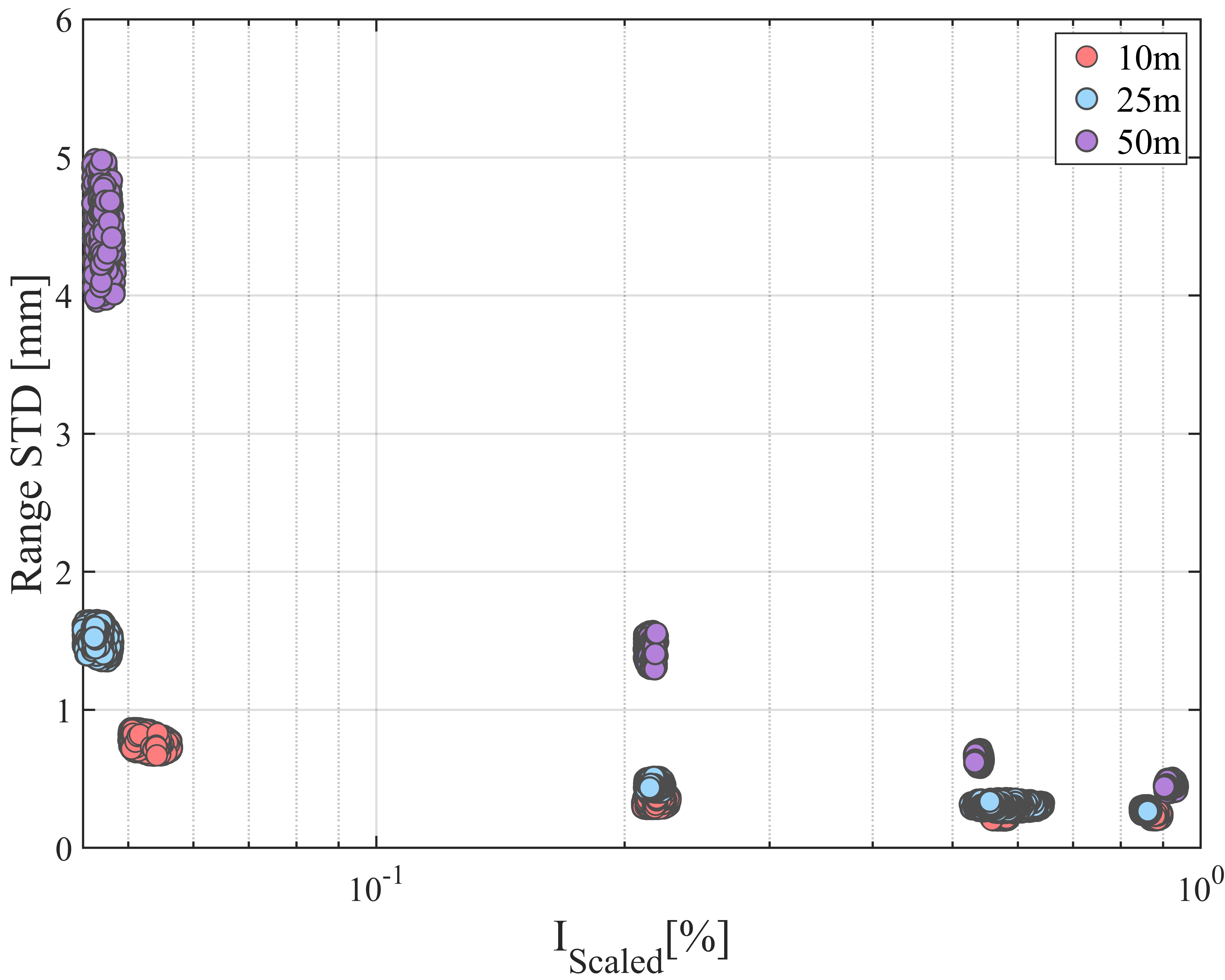}
\captionof{figure}{Exemplary relationship for Z+F~Imager~5016A using scaled intensity values without performing respective corrections, acquired at 10~m (orange), 25~m (blue), and 50~m (purple) for 1093.37~kHz scanning rate.}
\label{fig: ZF+Imager_10m_25m_50m_preprocessed_data sets_scaled_Intensity}
\end{center}
\vspace{1em}

To solve the limitations of using scaled intensities for range variance estimation, as mentioned in Section \ref{Scaled Intensity Values in Range Variance Estimation}, this section introduces a new procedure to overcome the distance-dependent effect observed in scaled intensity values (\autoref{fig: ZF+Imager_10m_25m_50m_preprocessed_data sets_scaled_Intensity}). The aim of this procedure is to calibrate the mean of the scaled intensity values so that a generalized intensity-based range variance model can be estimated, as illustrated in \autoref{fig: aim of the generalized model}.

\begin{center}
\includegraphics[width=\linewidth]{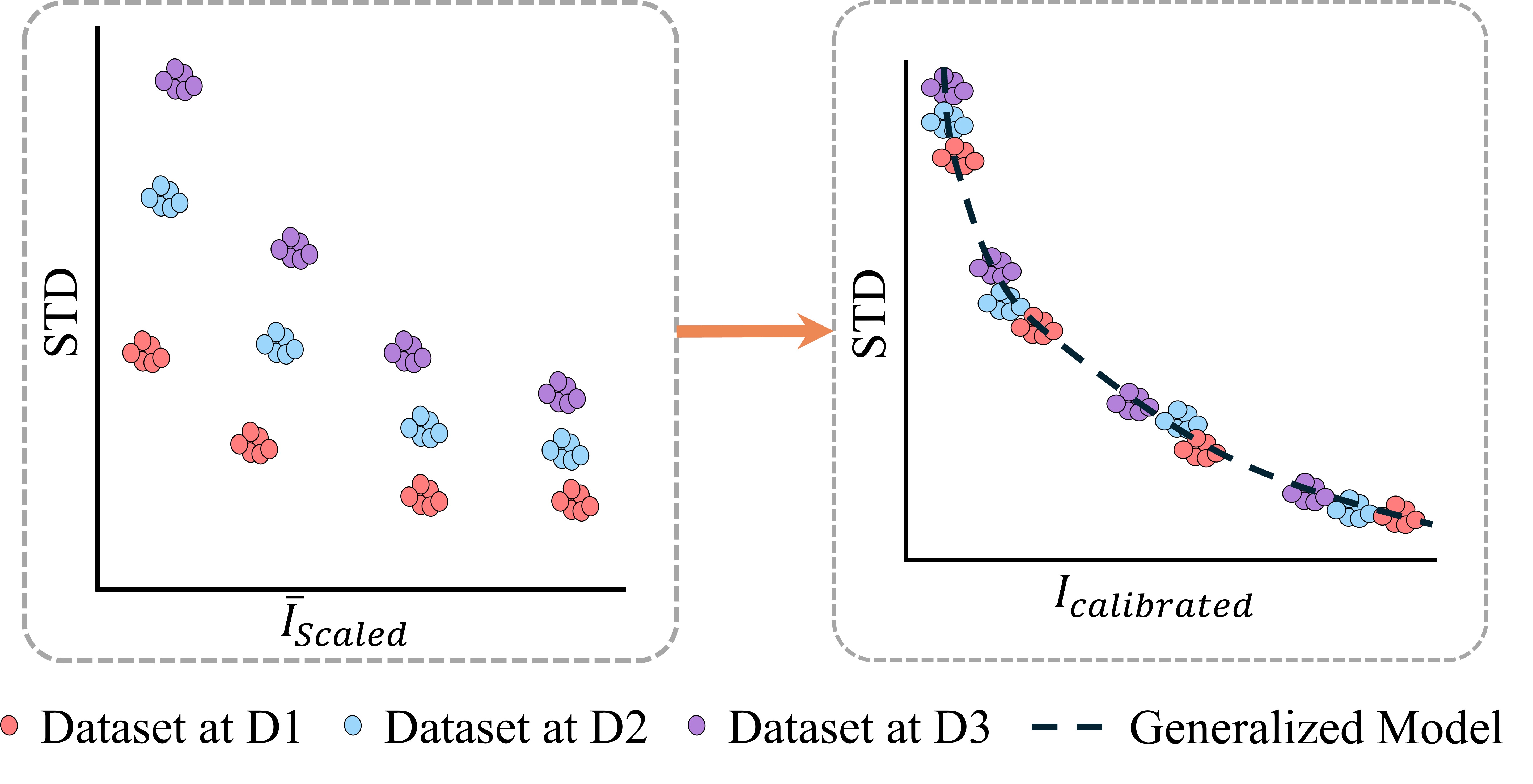}
\captionof{figure}{Illustration of the proposed calibration procedure for scaled intensity values. The left plot shows the relationship between range uncertainty and the mean scaled intensity values across all vertical ticks at three different distances ($D1 < D2 < D3$). The right plot presents the generalized model obtained after calibrating the intensities at different distances.}
\label{fig: aim of the generalized model}
\end{center}

\vspace{1em}

This calibration procedure of the intensity is based on the range between the scanner position and the scanned object.
\begin{equation}
    I_{\text{calib}} = \frac{\bar{I} \cdot r_{\text{ref}}}{\bar{r}^2},
\label{eq:calibrated_intensity}
\end{equation}
according to the decrease of intensity with squared range, known by the laser radar equation given in Eq. (\ref{eq:radar_range_equation}) \cite{jelalian_laser_1992}.

Therefore, we estimate the mean of the range $\bar{r}$ at each vertical tick using Eq. (\ref{eq:arithmetic mean}) as an extension of the workflow in \autoref{fig:Methodology_workflow}. Additionally, we define a reference range $r_{\text{ref}}$. This value may represent the average distance between the scanner and the scanned object. However, experimental results later revealed that this assumption does not hold consistently across different scanners. Therefore, the value can be defined by the user and adjusted according to the measurement range, as close-range data require a different reference value compared to medium- or long-range data. Since the scaling function is manufacturer-dependent, the reference range value also changes depending on the scanner used. This fact will be evaluated in Section \ref{Scaled Intensity _results}.



\subsection{Empirical estimation of the range variance model}
\label{Workflow for model fitting}

After data pre-processing and potential intensity calibration, the next step is to estimate the intensity-based range variance model by determining the model parameters $a$, $b$, and $c$, as defined in Eq. (\ref{eq:Intensity_based_range_variance_model}). To do so, the preprocessed mean intensity values for each vertical tick, $\bar{I}_j$, along with the corresponding range standard deviations, $\sigma_{r_j}$, are used in a parametric least-squares adjustment \citep{Mikhail1976}. This procedure can be applied to the means of both raw and scaled intensity values according to Figure \ref{fig:Methodology_workflow}. 

The model estimated from raw intensities is expected to be a single intensity-based range variance model valid across close, medium, and long ranges. However, when scaled intensities are used as input in the workflow proposed in \autoref{fig:Methodology_workflow}, multiple models arise that vary with the distance to the scanned object. Therefore, the generalized intensity-based range variance model is estimated between the mean of the calibrated intensity values $\bar{I}_j^{calib}$ and the corresponding range standard deviations to overcome the distance dependency.

\section{Experimental setup}
\label{Experimental Setup}

TLS manufacturers either offer to directly export the raw intensity values or they scale them using confidential correction functions before data export. Sometimes both options exist. In this paper, we use scanners that provide both raw and scaled intensity values like Z+F~Imager~5016A \citep{ZF_Imager_5016A}, as well as those that merely provide scaled intensity values like  Leica~ScanStation~P50 \citep{Leica_ScanStation_P50_2025}. Both scanners support panoramic (3D) and profile (2D) scanning modes for various scanning rates. In order to estimate the intensity-based range variance model, we adopt the 2D method mentioned in Section \ref{Methodology}.

As estimating the range standard deviations requires data redundancy, the measurement concept employs multi-profile scanning of a surface at close, medium, and long ranges for all available scanning rates. Consequently, we can cover a large amount of range variations and a large ntensity spectrum. Based on this, we introduce the measurement concept to estimate the model. 

The experiment is divided into two phases. The first phase aims to estimate the intensity-based range variance model in a controlled laboratory environment using raw and scaled intensity values (Section \ref{Data collection with spectralon board}), while the second phase seeks to estimate and evaluate the model's performance under real-world scenarios (Section \ref{Data collection for Model validation}). 

\subsection{Data collection with spectralon board (laboratory)}
\label{Data collection with spectralon board}

\paragraph*{Raw intensity values}
Since our goal is to estimate an intensity-based range variance model that captures both the full range of distances and the intensity spectrum, the experimental setup must be tailored to reflect these variations. Here, the Z+F~Imager~5016A was investigated, supporting both panoramic and profile scanning modes \citep{ZF_Imager_5016A}. Also, it provides for these two modes the same scanning settings, including scanning rates, rotation speeds, and point spacing, which can be configured by setting the resolution and the scanning quality. All accessible scanning settings were applied for our experiments.In this phase, the scanner remained
\begin{center}
\includegraphics[width=\linewidth]{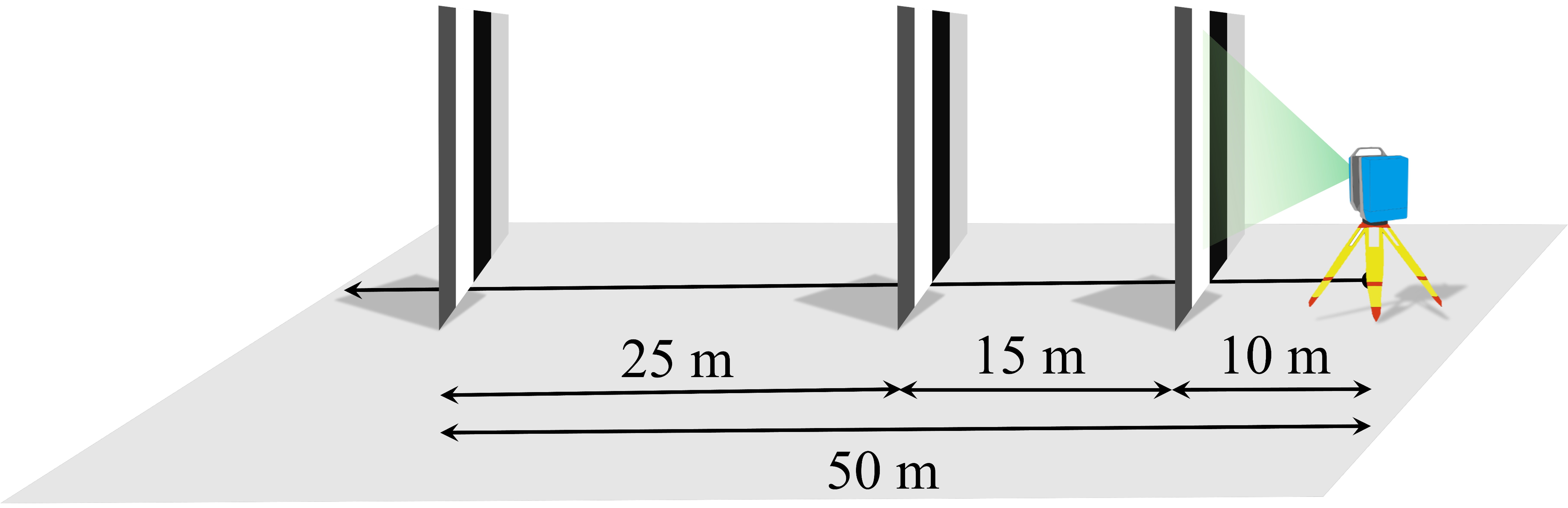}
\captionof{figure}{Measurement setup using Spectralon Board in profile mode at 10~m, 25~m, and 50~m distances; observed by Z+F~Imager~5016A.}
\label{fig:Lab_setup_zfImager5016A_spectralon}
\end{center}
\vspace{1em}
 stationary, while a Spectralon board ($1 \mathrm{m} \times 1 \mathrm{m}$, with colors/reflectance values dark gray/$\sim20\%$, white/$\sim90\%$, black/$\sim4.3\%$, light gray/$\sim53\%$) was positioned at three distances (10~m, 25~m, and 50~m) representing close, medium, and long range, as shown in \autoref{fig:Lab_setup_zfImager5016A_spectralon}. For each color of the board, more than 3000 profile lines were recorded to ensure a sufficient number of points for each target.

\paragraph*{Scaled intensity values}
The  Leica~ScanStation~P50 was employed. The drawback of using the profile mode in the  Leica~ScanStation~P50 is that not all scanning configurations are shared between the 2D mode and the panoramic mode. Thus, only the highest scanning configuration was applied. Unlike the Z+F~Imager~5016A, the scanning settings of this instrument are not defined by scanning rates but by specific resolution parameters, corresponding to a point spacing of 3.1~mm at 10~m, a rotation speed of 49.7 Hz, and a sampling rate of 1000 kPts/s.
%
\begin{center}
\includegraphics[width=\linewidth]{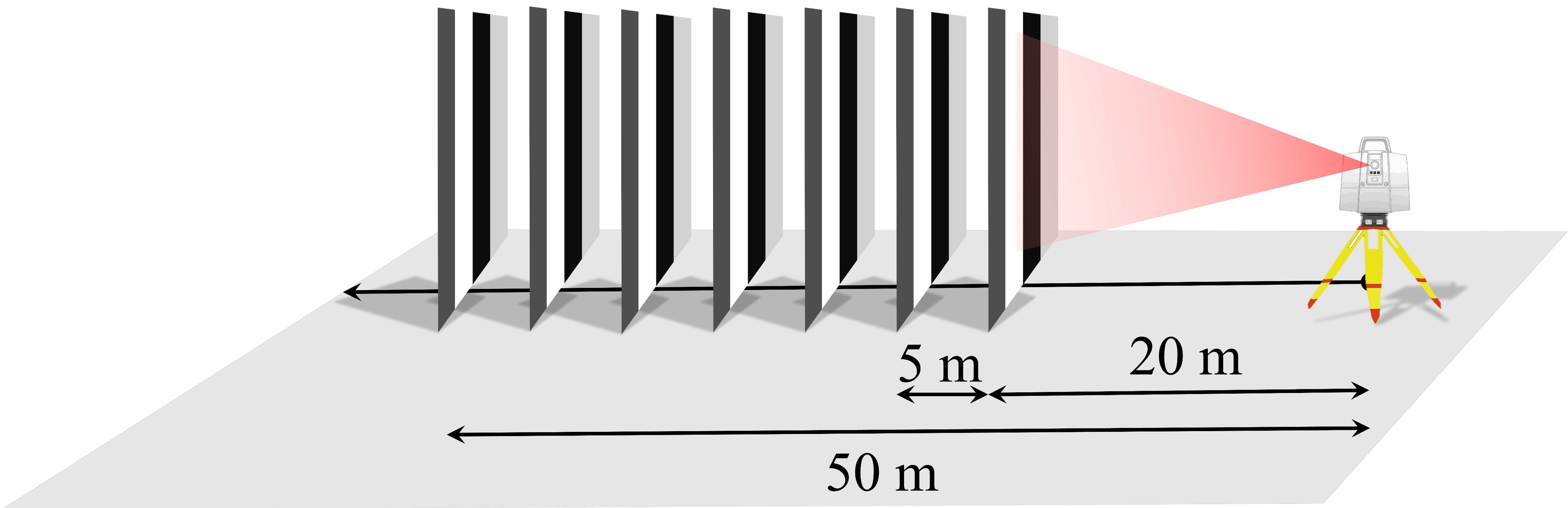}
\captionof{figure}{Measurement configuration using Spectralon Board observed by Leica~ScanStation~P50 in profile mode from 20~m to 50~m Distance with a 5~m interval.}
\label{fig:Lab_setup_LeicaP50_spectralon}
\end{center}
\vspace{1em}
The Spectralon board was placed at 20~m from the scanner and moved up to 50~m in a 5~m interval, as shown in Figure \ref{fig:Lab_setup_LeicaP50_spectralon}. This allowed us to investigate the effect of the scaling function on the estimation of range uncertainty. 

\subsection{Data collection for model evaluation}
\label{Data collection for Model validation}

\begin{figure}
   \centering
    \includegraphics[width=\linewidth]{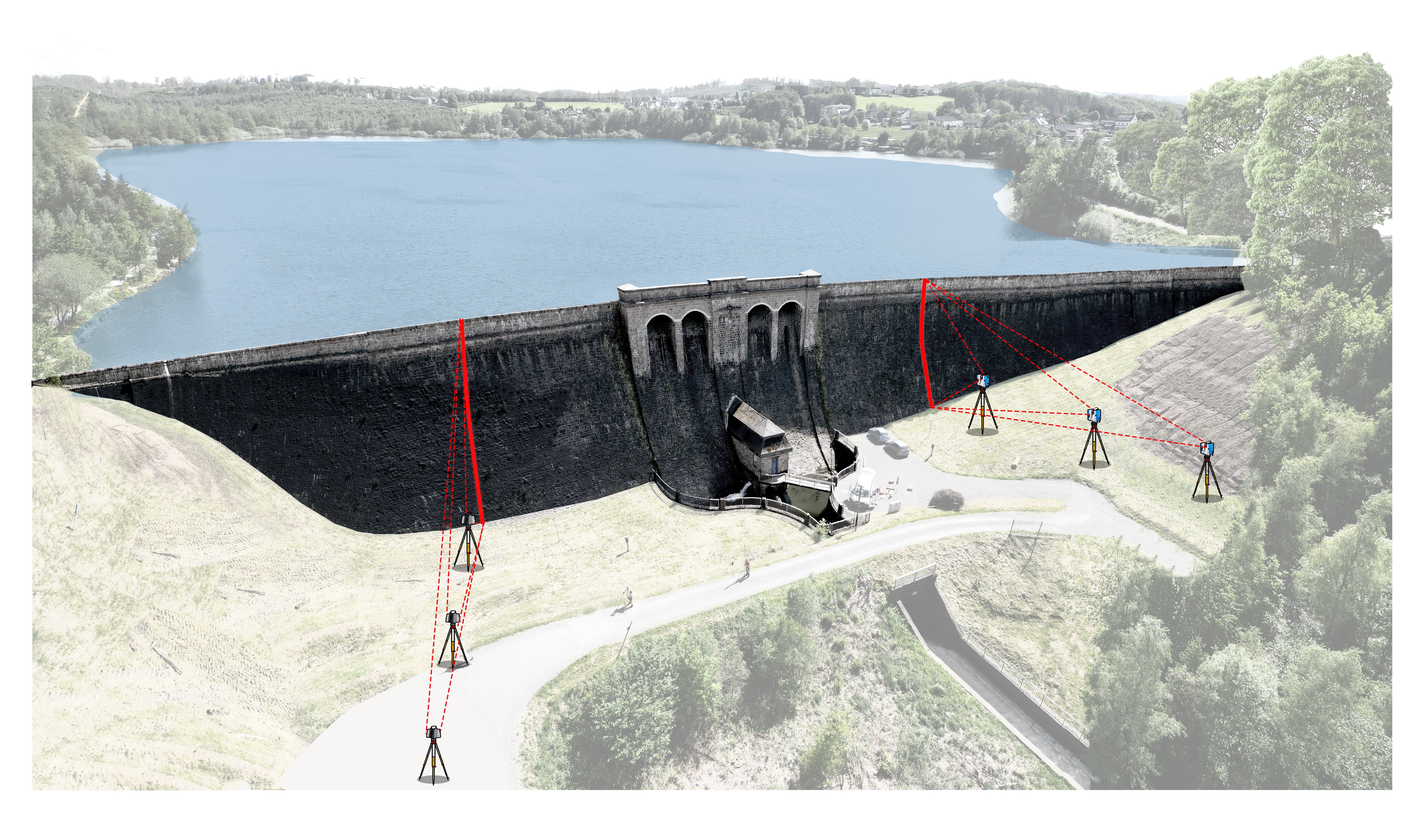}
    \caption{Measuring configuration at the Brucher Water Dam, where two scanning lines were placed on each side at distances of 10~m, 20~m, and 30~m, where the scanners used are Z+F~Imager~5016A and  Leica~ScanStation~P50.}
    \label{fig:waterdam_data_validation}
\end{figure}

\begin{figure}
   \centering
    \includegraphics[width=\linewidth]{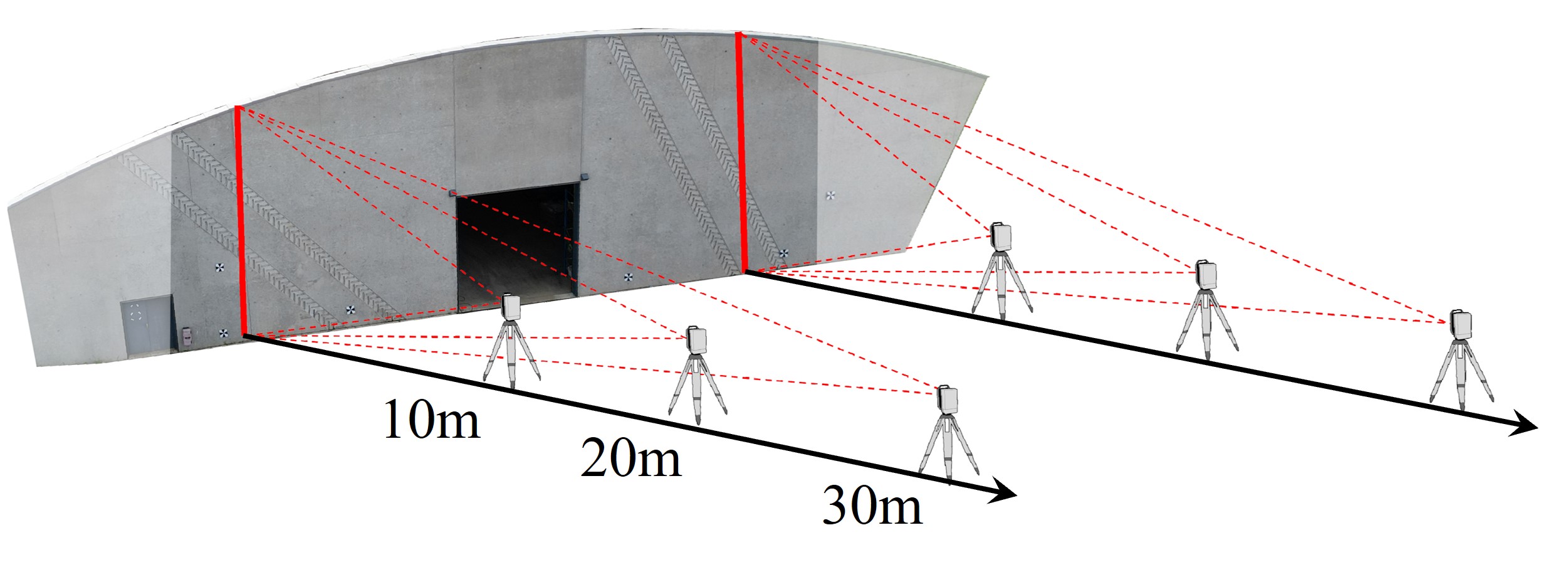}
    \caption{Measurement setup at the Bonn Reference Wall, where two scanning lines were placed on each side at distances of 10~m, 20~m, and 30~m, as indicated by the white points. The measurements were carried out using the Z+F~Imager~5016A.}
    \label{fig: Bonn Reference Wall_data_validation}
\end{figure}

In the second phase, the objective is to evaluate the proposed workflow under typical environmental conditions, involving various surface materials and geometries: A single profile scanning line was captured for both a Metallic Art Sculpture (\autoref{fig:Art_sculupture_data_validation}) and a concrete wall at the Pinakothek der Moderne (Gallery of Modern Art, Munich, \autoref{fig:Pinakothek der Moderne (Gallery of Modern Art, Munich)_data_validation}). The scanning was carried out using the Z+F~Imager~5016A at a scanning rate of 546.495~kHz and distances of 10~m, 25~m, and 30~m.
Additionally, to examine the effect of surface complexity, scans were acquired on each side of the Brucher Water Dam (\autoref{fig:waterdam_data_validation}) and the Bonn Reference Wall (\autoref{fig: Bonn Reference Wall_data_validation}) at distances of 10~m, 20~m, and 30~m. The scanning at the water dam was performed using both the Z+F~Imager~5016A and the Leica~ScanStation~P50, 
\begin{center}
\includegraphics[width=\linewidth]{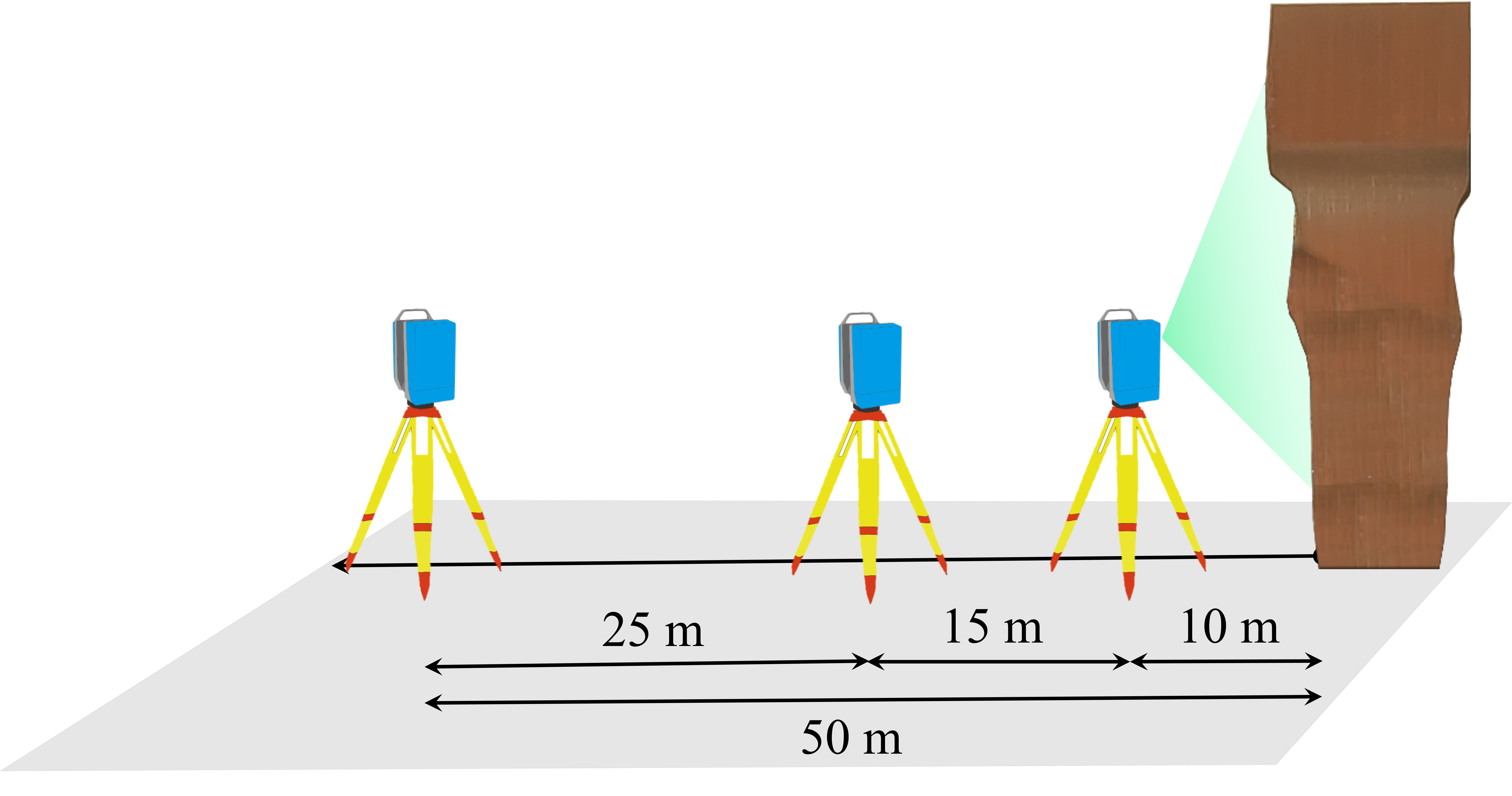}
\captionof{figure}{Measuring setup using a Metallic Art Sculpture observed by Z+F~Imager~5016A at 10~m, 25~m, and 50~m distances in profile mode.}
\label{fig:Art_sculupture_data_validation}
\end{center}

\begin{center}
\includegraphics[width=0.4\linewidth]{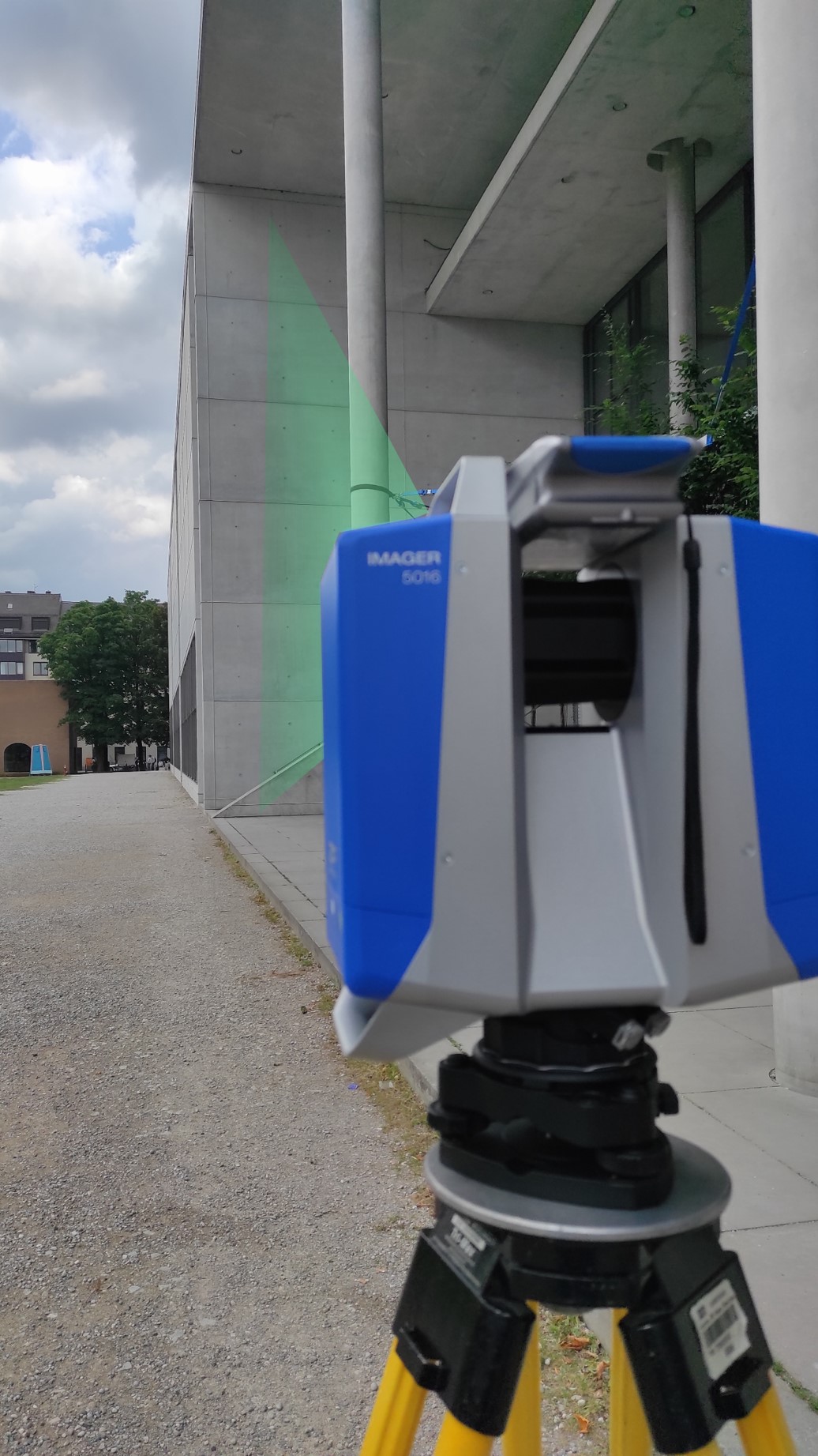}
\captionof{figure}{The Pinakothek der Moderne validation data set acquisition using Z+F~Imager~5016A at 10~m, 25~m, and 30~m applying profile mode.}
\label{fig:Pinakothek der Moderne (Gallery of Modern Art, Munich)_data_validation}
\end{center}
\vspace{1em}
whereas only the Z+F~Imager~5016A was employed at the Bonn Reference Wall. For the Z+F~Imager~5016A, a scanning rate of 1093.37~kHz was used with point spacing of 3.1~mm at 10~m , while the Leica~ScanStation~P50 operated at the same point spacing.


\section{Results and model evaluation}

\label{Results and Model Validation}

Following data acquisition, the workflow outlined in Section~\ref{Methodology} is applied to the collected data. This section presents the results for both raw and scaled intensity data sets. In Section~\ref{Raw Intensity _results}, it covers the intensity-based range variance models for different scanning rates using laboratory data sets collected with the Z+F~Imager~5016A and estimated from the raw intensity values, followed by evaluating these models using real-world data sets. Then, the possibility of estimating the model directly from real-world object data sets is introduced. 
\twocolumn[{
\begin{center}
\captionof{table}{Scanning settings and intensity-based range variance model parameters of the Z+F~Imager~5016A for 2D profile mode at various resolutions, quality modes, and scanning rates.}
\label{tab:z+fImager5016A_scanning_settings_and_parameters}
\begin{tabular}{llllllll}
\starttabularbody
Resolution   & Quality  & Scanning rate [kHz] & Mirror speed [rps] & Point spacing & $a$ [$\frac{\text{mm}}{\text{INC}}$] & $b$ [--] & $c$ [mm] \\ \midrule

Low          & low      & 136.671   & 54 & 25.1~mm @ 10m & 29853  & -1.02 & 0.08 \\
             & balanced & 68.312    & 27 & 25.1~mm @ 10m & 35162  & -1.07 & 0.08 \\
             & high     & 34.132    & 13 & 25.1~mm @ 10m & 37371  & -1.11 & 0.07 \\

Middle       & low      & 273.343   & 54 & 12.6~mm @ 10m & 44509  & -1.02 & 0.09 \\
             & normal   & 136.624   & 27 & 12.6~mm @ 10m & 39138  & -1.04 & 0.09 \\
             & high     & 68.264    & 13 & 12.6~mm @ 10m & 31742  & -1.06 & 0.08 \\
             & premium  & 34.085    & 6  & 12.6~mm @ 10m & 30698  & -1.09 & 0.07 \\

High         & low      & 546.685   & 54 & 6.3~mm @ 10m  & 59140  & -1.01 & 0.11 \\
             & normal   & 273.248   & 27 & 6.3~mm @ 10m  & 41432  & -1.01 & 0.09 \\
             & high     & 136.529   & 13 & 6.3~mm @ 10m  & 42183  & -1.05 & 0.09 \\
             & premium  & 68.169    & 6  & 6.3~mm @ 10m  & 35138  & -1.07 & 0.08 \\

Super High   & low      & 1093.370  & 54 & 3.1~mm @ 10m  & 100195 & -1.03 & 0.21 \\
             & normal   & 546.495   & 27 & 3.1~mm @ 10m  & 53076  & -1.00 & 0.10 \\
             & high     & 273.058   & 13 & 3.1~mm @ 10m  & 45268  & -1.02 & 0.10 \\
             & premium  & 136.529   & 6  & 3.1~mm @ 10m  & 42183  & -1.05 & 0.09 \\

Ultra High   & normal   & 1092.990  & 27 & 1.6~mm @ 10m  & 103106 & -1.03 & 0.21 \\
             & high     & 546.495   & 13 & 1.6~mm @ 10m  & 53076  & -1.00 & 0.10 \\
             & premium  & 272.678   & 6  & 1.6~mm @ 10m  & 62382  & -1.05 & 0.10 \\

Extremely High & high   & 1091.850  & 10 & 0.6~mm @ 10m  & 92321  & -1.02 & 0.19 \\
             & premium  & 544.975   & 5  & 0.6~mm @ 10m  & 56230  & -1.01 & 0.11 \\

\end{tabular}
\end{center}
\vspace{1em}
}]
In Section~\ref{Scaled Intensity _results}, the estimation of the intensity-based range variance model is performed using the scaled intensity data sets collected with both the Leica~ScanStation~P50 and the Z+F~Imager~5016A in two approaches: one using the scaled intensity values directly and the other after calibration. Afterwards, each approach is evaluated using the Brucher Water Dam data sets acquired at 20~m and 30~m.
\subsection{Results using raw intensity} 
\label{Raw Intensity _results}

\subsubsection{Model estimation based on laboratory data set}

The raw intensity-based range variance models for the Z+F~Imager~5016A are estimated for all available scanning rates. Numerical results are given in Table \ref{tab:z+fImager5016A_scanning_settings_and_parameters} and in \autoref{fig: ZF_all_Intensity based range variance models_raw_intensity}: The relationship between the mean intensity values and range uncertainties demonstrates that for all scanning rates, targets with low intensities observed from long ranges exhibit high range uncertainties, whereas targets with high intensities and short ranges show lower uncertainties. Thus, for each scanning rate, a single model can represent the intensity-based range variance across close, medium, and long ranges. 

As the scanning rate increases, the range noise also increases. For example, the range uncertainty of a black target observed at long range with a scanning rate of 34.085~kHz is lower than that of the same target scanned at 1093.37 kHz. In Table \ref{tab:z+fImager5016A_scanning_settings_and_parameters}, it can be observed that several scanning rates are nearly identical. This is also reflected in the estimated parameters 
$a$, $b$, and $c$, which show only minor variations across these similar scanning configurations, for example, between the scanning rates of 1093.370~kHz and 1092.990 kHz.

This general outcome is the expected behavior according to literature \citep{schill_intensity-based_2024,wujanz_intensity-based_2017,heinz_strategy_2018,jost_using_2020}. Individual estimates of parameters $a,b,c$ cannot be evaluated using literature since they have not been determined for the Z+F~Imager~5016A yet.

\begin{center}
\includegraphics[width=\linewidth]{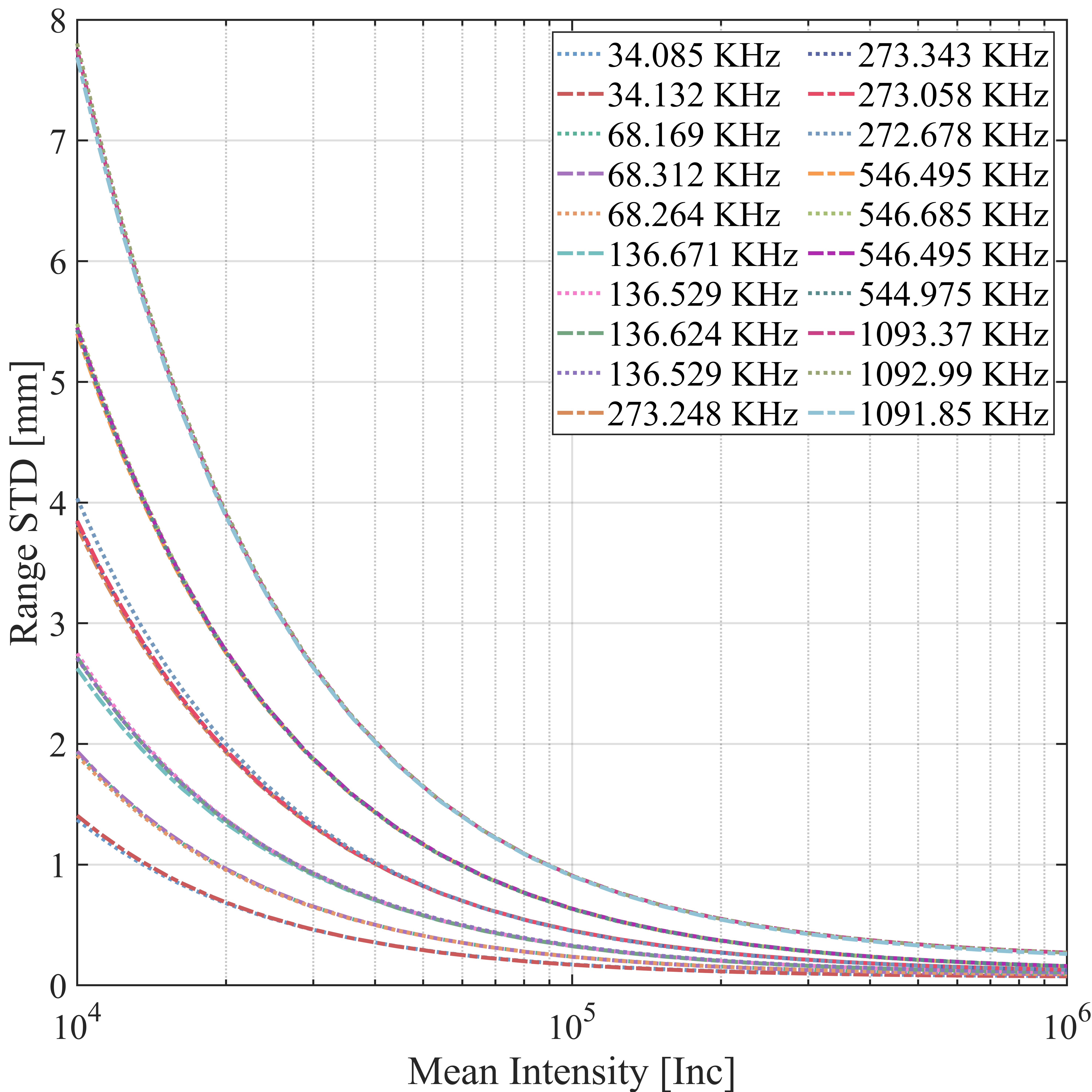}
\captionof{figure}{Raw intensity-based range variance models of the Z+F~Imager~5016A across all available scanning rates after parameter estimation, arranged from the lowest to the highest scanning rates.}
\label{fig: ZF_all_Intensity based range variance models_raw_intensity}
\end{center}
\vspace{4em}


\subsubsection{Model evaluated based on real-world data}\label{Model evaluated based on real-world data}
After the intensity-based range variance models were estimated under laboratory conditions, two validation sequences took place. A cross-validation was first performed by using two scanning rates. The Pinakothek Wall and the Metallic Art Sculpture were both scanned at a scanning rate of 546.495 kHz, while the Bonn Reference Wall and the Brucher Water Dam were scanned at 1093.37 kHz.

Examining the preprocessed data of each target and plotting the mean intensities and the corresponding range standard deviations for all four targets against the intensity-based range variance models estimated based on the laboratory data set, as shown in \cref{fig:Modelvalidation_model_comparison_Pinakothek_wall_and_metallic_art_sculpture,fig:Modelvalidation_model_comparison_the_Brucher_water_dam_and_Bonn_reference_wall}, it can be observed that the data distribution follows a trend similar to that of the estimated model. However, in \cref{fig:Modelvalidation_model_comparison_the_Brucher_water_dam_and_Bonn_reference_wall}, on the left side where the Brucher Water Dam data set appears, it slightly deviates from the model due to its concave geometry and rough surface structure.

By estimating the root mean square error (RMSE) of the residuals at each of the four targets, we find that all values lie within the sub-millimeter range: 0.01~mm for the Pinakothek Wall, 0.02~mm for the Metallic Art Sculpture, 0.03~mm for the Bonn Reference Wall, and 0.07~mm for the Brucher Water Dam, with corresponding maximum residuals of 0.05~mm, 0.13~mm, 0.10~mm, and 0.18~mm, respectively. The variation in the maximum residuals arises from the differing surface characteristics of the observed targets, spanning from smooth planar structures in the Bonn wall and the Pinakothek to the moderately curved metallic surface of the art sculpture and the geometrically intricate and rough structure of the water dam. These results motivate an assessment of whether individually fitted models could be derived for each data set.

\begin{center}
\includegraphics[width=\linewidth]{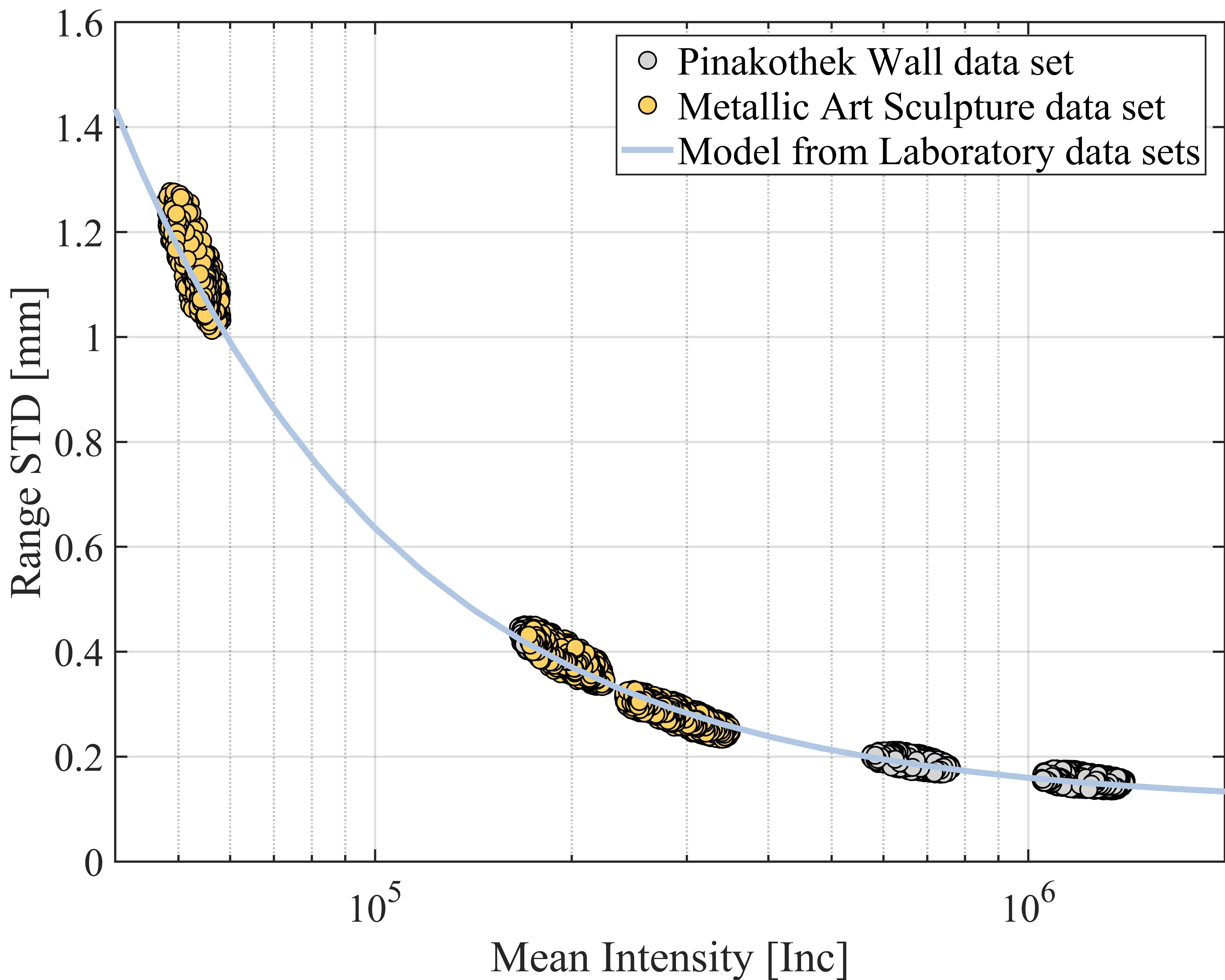}
\captionof{figure}{The Pinakothek Wall and Metallic Art Sculpture data sets compared with the intensity-based range variance model at a scanning rate of 546.495 kHz.}
\label{fig:Modelvalidation_model_comparison_Pinakothek_wall_and_metallic_art_sculpture}
\end{center}

\begin{center}
\includegraphics[width=\linewidth]{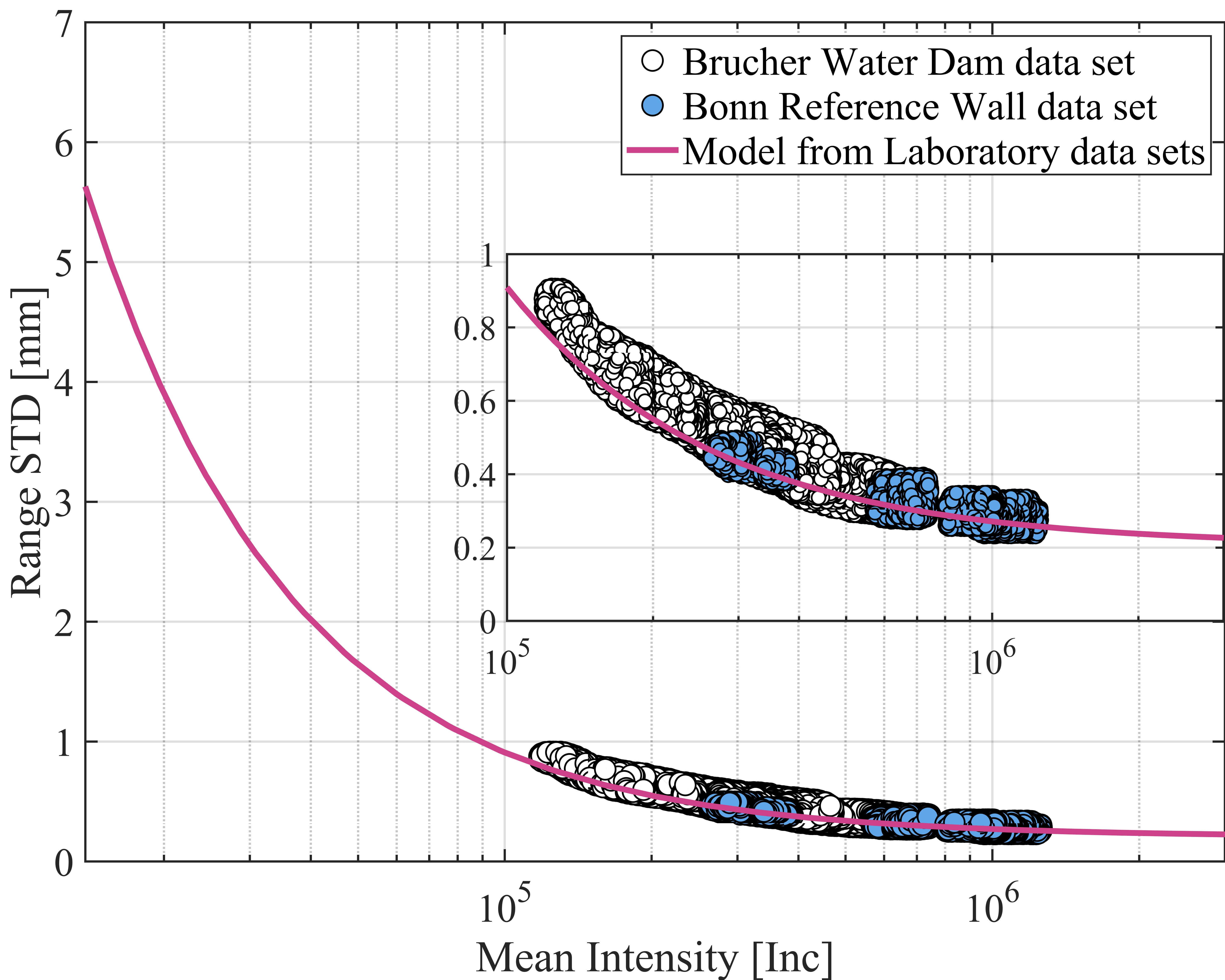}
\captionof{figure}{Comparison of the behavior of the Brucher Water Dam and Bonn Reference Wall data sets with the intensity-based range variance model at the same scanning rate of 1093.37 kHz.}
\label{fig:Modelvalidation_model_comparison_the_Brucher_water_dam_and_Bonn_reference_wall}
\end{center}

\subsubsection{Model estimation based on real-world data sets}
Applying the parametric least-squares adjustment of Section \ref{Workflow for model fitting} to the real-world data sets, the parameters for each function are estimated (\autoref{tab:intensity_based_range_variance_parameters_from_natural_targets}). Comparing the model parameters derived from the Pinakothek Wall and the Metallic Art Sculpture data sets with those of the intensity-based range variance model at a scanning rate of 546.495~kHz (\autoref{tab:z+fImager5016A_scanning_settings_and_parameters}), differences in the parameters $a$, $b$, and $c$ are observed with maximal ranges of 44018~mm/INC, 0.148, and 0.09~mm, respectively. These variations indicate that $a$ is the most influenced by the surface properties of the observed targets, while $b$, and $c$ remain relatively stable, reflecting similar model behavior across the different data sets. 

\begin{center}
\captionof{table}{Estimated parameters for the intensity-based range variance models derived from Pinakothek Wall, the Metallic Art Sculpture acquired at 546.495 kHz, Bonn Reference Wall, and Brucher Water Dam data sets at 1093.37 kHz.}
\label{tab:intensity_based_range_variance_parameters_from_natural_targets}
\begin{tabular}{llll}
\starttabularbody
Evaluation data sets          & $a$ [$\frac{\text{mm}}{\text{INC}}$] & $b$ [--] & $c$ [mm] \\ \midrule
Pinakothek Wall              & 85437      & -1.038	& 0.11            \\
Metallic Art Sculpture       & 11880	  & -0.852	& 0.02             \\
Bonn Reference Wall          & 4386	      & -0.775	& 0.19              \\
Brucher Water Dam            & 18516	  & -0.872	& 0.18               \\
\end{tabular}
\end{center}
\vspace{1em}

For the model estimated from the Pinakothek Wall data set, as shown in \autoref{fig:Intensity-based range variance models derived from the Spectralon board and the Pinakothek Wall datasets}, the data are concentrated in the flat region of the model curve, where intensity values are relatively high and range noise is low. This results in a steeper decay, reflected in the more negative value of $b$, and a larger scaling factor $a$. On the other hand, the model estimated from the Metallic Art Sculpture data set, as illustrated in \autoref{fig:Intensity-based range variance models derived from the Spectralon board and the Metallic Art Sculpture datasets}, is based on data dominated by lower intensities. As a result, this model has a smaller scaling factor $a$, a less negative exponent $b$, and a lower noise floor $c$.

\begin{center}
\includegraphics[width=\linewidth]{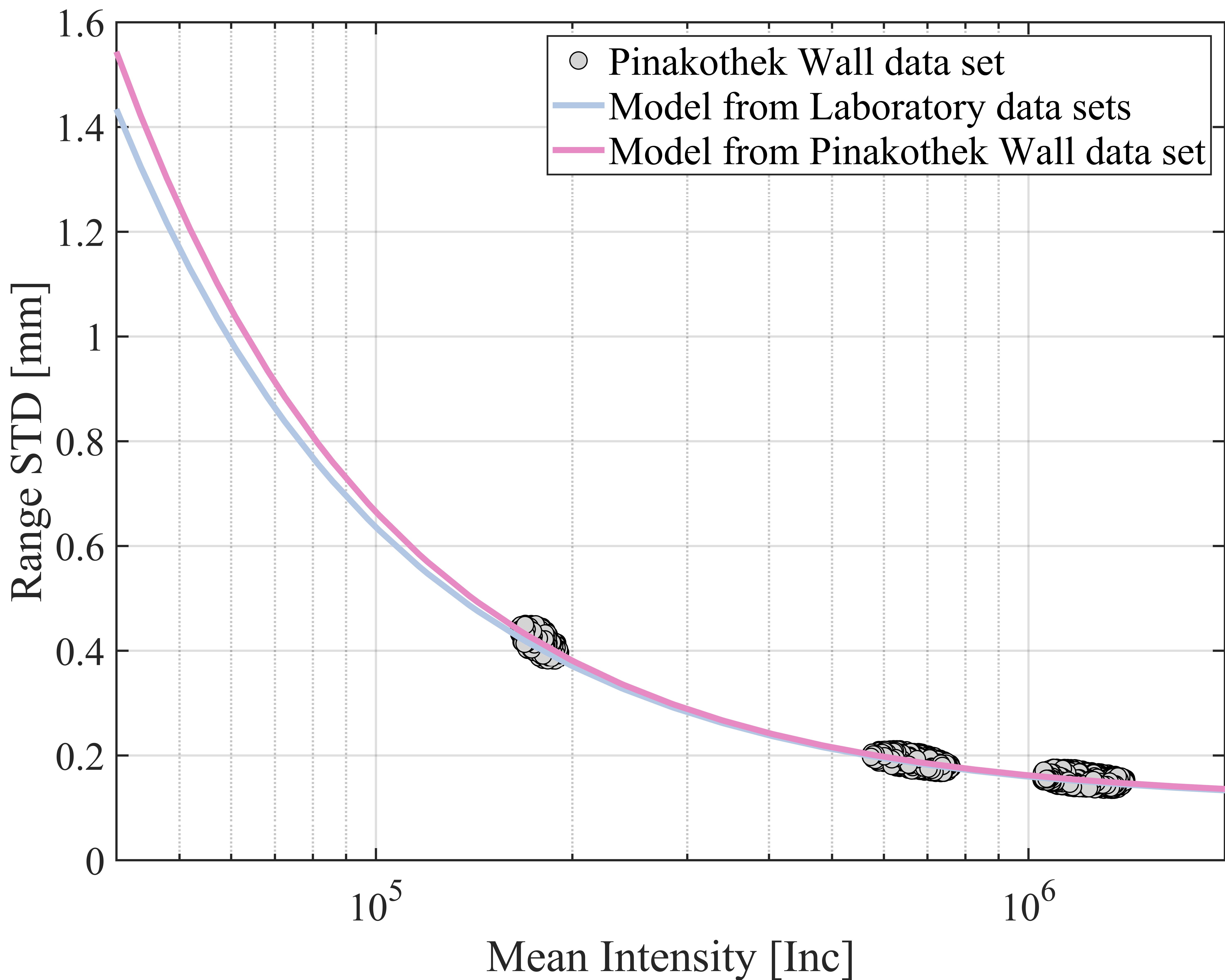}
\captionof{figure}{Comparison of intensity-based range variance models derived from the Spectralon board and the Pinakothek Wall data set, acquired at a scanning rate of 546.495 kHz. The model based on Spectralon observations is illustrated by a solid blue curve, whereas the Pinakothek Wall model is represented by a pink curve.}
\label{fig:Intensity-based range variance models derived from the Spectralon board and the Pinakothek Wall datasets}
\end{center}

\autoref{fig:Modelvalidation_model_comparison_the_Brucher_water_dam_and_Bonn_reference_wall_spectralon_full_spectrum} presents the model curves derived from the laboratory-based estimation at a scanning rate of 1093.37~kHz (\autoref{tab:z+fImager5016A_scanning_settings_and_parameters}) along with those estimated from the Brucher Water Dam and Bonn Reference Wall data sets (\autoref{tab:intensity_based_range_variance_parameters_from_natural_targets}) across the entire intensity spectrum. A zoomed-in plot displays the same models but focused on the limited intensity spectrum covered by the Brucher Water Dam and Bonn Reference Wall data sets. Within this limited range, the differences between the models remain below 0.2~mm, with the largest deviations occurring at lower intensity values. However, when the full intensity spectrum is considered, as shown in \autoref{fig:Modelvalidation_model_comparison_the_Brucher_water_dam_and_Bonn_reference_wall_spectralon_full_spectrum}, the differences increase substantially, reaching up to approximately 3~mm at low intensities. This highlights a clear extrapolation effect caused by the absence of low-intensity observations in the evaluation data sets.


The variation between the models estimated from the Bonn Reference Wall and the water dam data sets, compared with the model fitted from the laboratory data set, is due to the differing surface characteristics of the targets.
\begin{center}
\includegraphics[width=\linewidth]{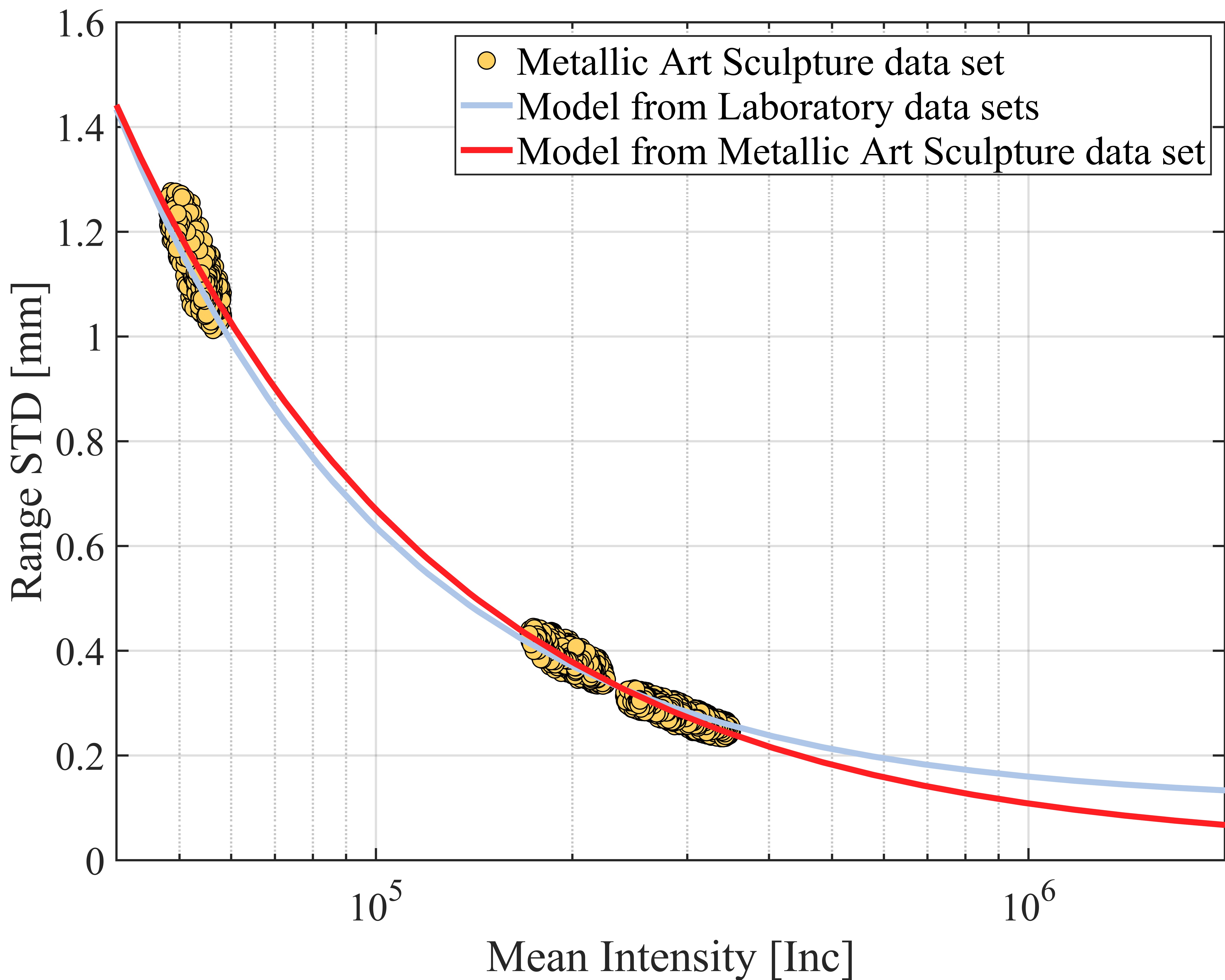}
\captionof{figure}{Comparison of intensity-based range variance models derived from the Spectralon board and the Metallic Art Sculpture data set, both acquired at a scanning rate of 546.495 kHz. The Spectralon-based model is depicted as a solid blue curve, while the Metallic Art Sculpture model is represented by a red curve.}
\label{fig:Intensity-based range variance models derived from the Spectralon board and the Metallic Art Sculpture datasets}
\end{center}
This also explains the differences in their parameters $a$, $b$, and $c$. The model based on the Brucher Water Dam data shows a lower scaling factor $a$ and a less steep decay exponent $b$, which reflects the concave geometry and the heterogeneous composition of its rocky surface that increase scattering and reduce the effective backscattered signal, thereby influencing the model parameters. In comparison, the model derived from the Bonn Reference Wall shows the lowest scaling factor $a$ and the shallowest decay rate (least negative $b$), despite the smooth and flat nature of the surface. The noise floor $c$ remains similar to that of the Spectralon-based model, indicating a consistent level of baseline system noise across all three data sets.

As further evaluation, we compare the models derived from the real world data sets with the laboratory-based models on the basis of RMSE and the maximum residuals. Here, we calculate residuals between the range standard deviations predicted by the Pinakothek Wall and Metallic Art Sculpture models and those from the laboratory model at 546.495 kHz. Similarly, we compare the Brucher Water Dam and Bonn Reference Wall data sets using the laboratory model at 1093.37 kHz. Subsequently, the RMSE and maximum absolute residuals were computed to quantify the discrepancies, as presented in \autoref{tab:validation_results_RMSE_residual_model_comparison}.

By comparing the RMSE and maximum residual values from the evaluation of the models estimated from the laboratory data sets using the Pinakothek Wall, Metallic Art Sculpture, Brucher Water Dam, and Bonn Reference Wall data sets mentioned in the previous section, it becomes evident that both metrics decrease when the models
\begin{center}
\includegraphics[width=\linewidth]{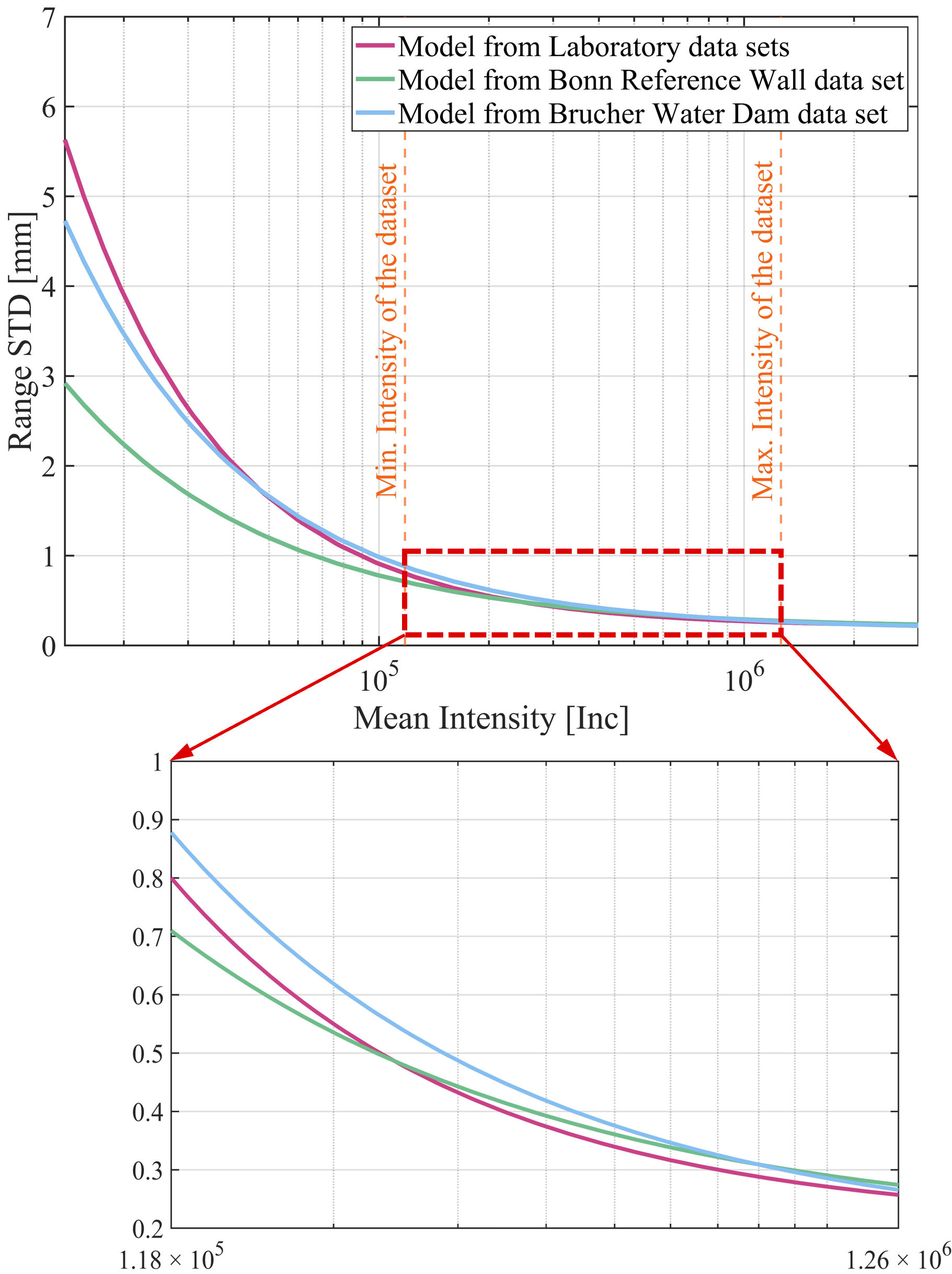}
\captionof{figure}{Intensity-based range variance models derived from the Spectralon board (pink solid line), Bonn Reference Wall (green solid line), and Brucher Water Dam (blue solid line) data sets, covering the full intensity spectrum. All data sets were acquired using the Z+F~Imager~5016A operating at a scanning rate of 1093.37 kHz. The zoomed-in plot shows the intensity range observed in the Bonn Reference Wall and Brucher Water Dam data sets.}
\label{fig:Modelvalidation_model_comparison_the_Brucher_water_dam_and_Bonn_reference_wall_spectralon_full_spectrum}
\end{center}

\begin{center}
\captionof{table}{RMSE and maximum absolute residual between the standard deviations predicted by the intensity-based range variance model fitted to the Spectralon board and those estimated from the four data sets.}
\label{tab:validation_results_RMSE_residual_model_comparison}
\begin{tabular}{lll}
\starttabularbody
Evaluation data sets          & RMSE [mm] & Max. Residuals [mm] \\ \midrule
Pinakothek Wall              & 0.01      & 0.01                \\
Metallic Art Sculpture       & 0.01      & 0.03                \\
Bonn Reference Wall          & 0.02      & 0.02                \\
Brucher Water Dam            & 0.06      & 0.08                \\
\end{tabular}
\end{center}
\vspace{1em}
are estimated from the data sets of their corresponding real-world targets. However, this observation holds only within the intensity range covered by each data set. For instance, in the Brucher Water Dam data, the maximum residual is reduced from 0.18~mm to 0.08~mm, illustrating the improvement achieved when using the model derived from the same target.

\begin{figure*}
    \centering
    {%
    \captionsetup[subfigure]{justification=centering}
    \begin{subfigure}[t]{0.48\textwidth}
        \centering
        \includegraphics[width=\linewidth]{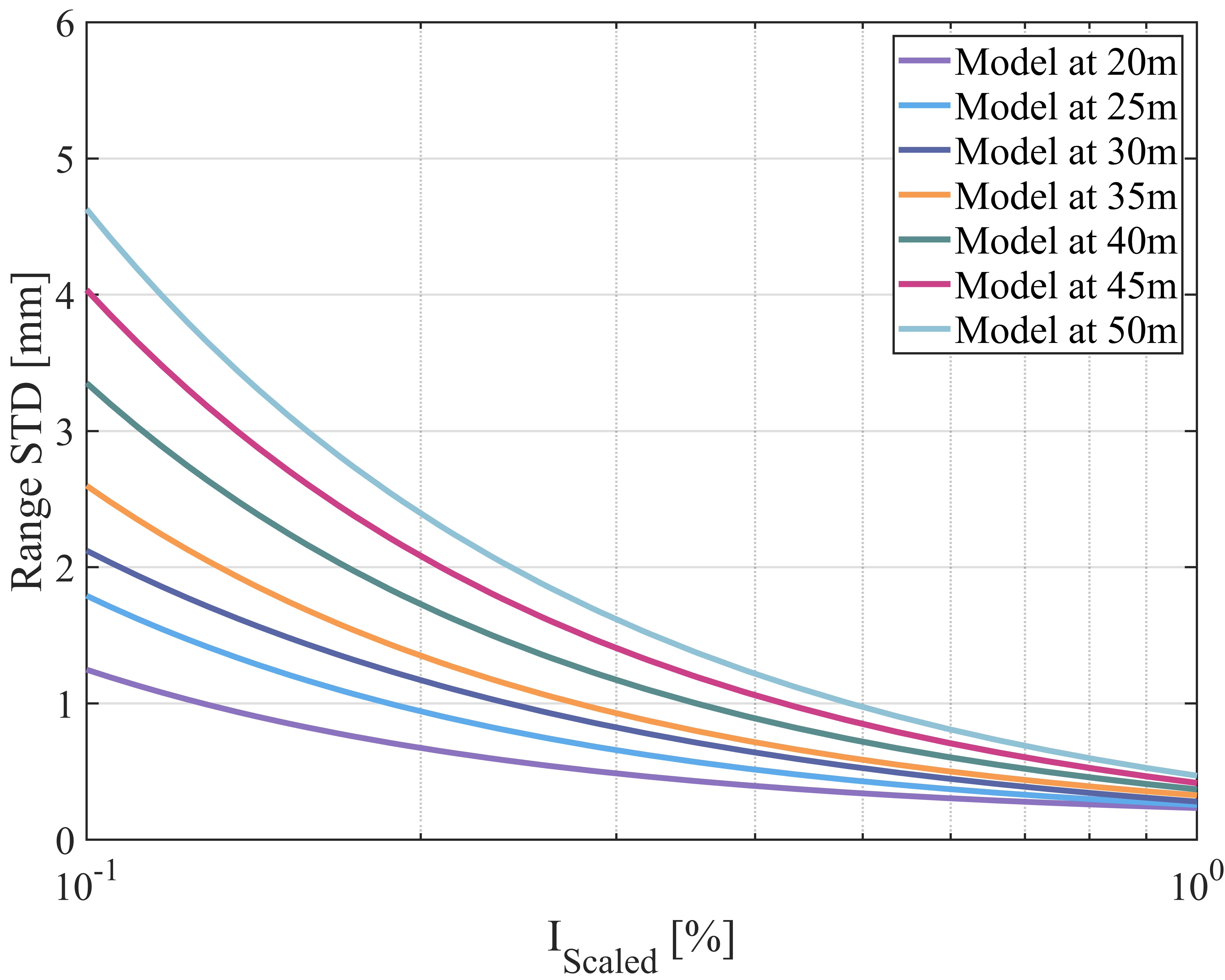}
         \caption{}
        \label{fig:Intensity-based range variance models from scaled intensity values_laboratory data set_Leica P50}
    \end{subfigure}
    \hfill
    \begin{subfigure}[t]{0.48\textwidth}
        \centering
        \includegraphics[width=\linewidth]{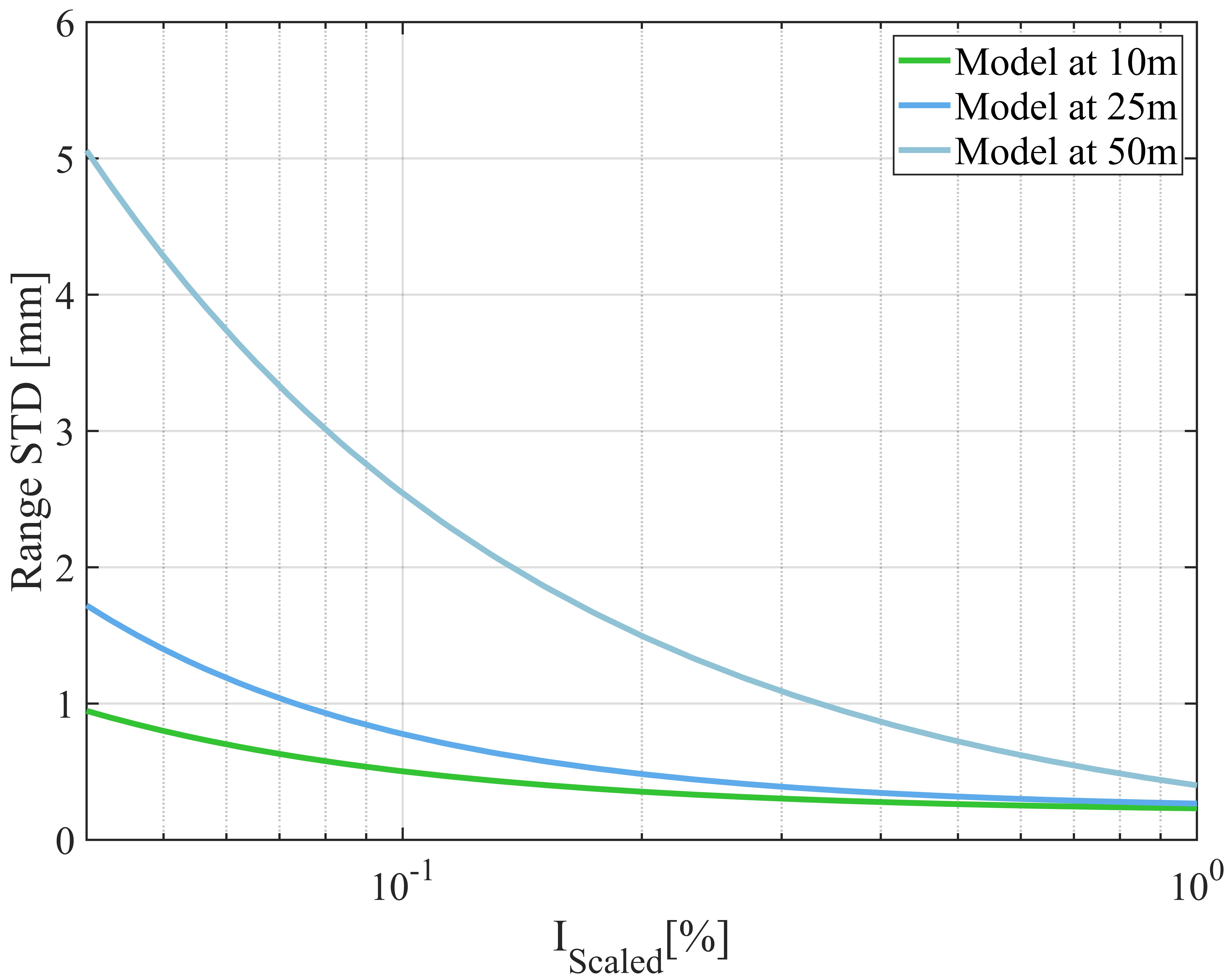}
        \caption{}
        \label{fig:Intensity-based range variance models from scaled intensity values_laboratory data set_Z+FImager5016A}
    \end{subfigure}
    }
    \caption{Intensity-based range variance models estimated from scaled intensity data acquired under laboratory conditions using a Spectralon board with a point spacing of 3.1~mm at 10~m. (a) Models derived from the Leica~ScanStation~P50 at distances between 20~m and 50~m in 5~m intervals. (b) Models derived from the Z+F~Imager~5016A using data sets acquired at 10~m, 25~m, and 50~m with a scanning rate of 1093.37 kHz.}
    \label{fig:model_comparison_scaled intensities_p50_ZF}
\end{figure*}

\subsubsection{Summarizing discussion}
The models estimated from the laboratory data sets show the expected behavior, where the range noise decreases as the signal strength increases. The scanning rates of the Z+F~Imager~5016A show noticeable differences in the absolute noise levels, confirming that higher scanning rates lead to higher range noise. These results emphasize the importance of the laboratory-based model as a reliable reference for evaluating the scanner’s performance under controlled conditions. When evaluated using real-world data sets, the laboratory-derived models fit well within the intensity range covered by the observations, demonstrating their applicability and transferability to practical measurement scenarios. The models estimated from the real-world data also work well within the range of the intensity spectrum used for their derivation; otherwise, extrapolation effects occur. Thus, the model could also be derived on-site but would only be usable for this specific range of intensities. Anyhow, this is an important finding, as it allows for an easy derivation and use of the model directly in real measurement scenarios. 


\subsection{Results using scaled intensity values} 
\label{Scaled Intensity _results}
\subsubsection{Model estimation without intensity calibration}


To demonstrate the effect of scaled intensity, the intensity-based range variance models are estimated from the laboratory data sets acquired with the Leica~ScanStation~P50 using a point spacing of 3.1~mm at 10~m and the Z+F~Imager~5016A with the same point spacing at a scanning rate of 1093.37~kHz while observing the Spectralon board, using the mean scaled intensity values at each distance and following the workflow illustrated in \autoref{fig:Methodology_workflow}. The results presented in \autoref{fig:model_comparison_scaled intensities_p50_ZF} illustrate the model estimation for both scanners. For the Leica~ScanStation~P50, an individual model is estimated for each distance starting from 20~m up to 50~m. For the Z+F~Imager~5016A, one model is estimated for 10~m, one for 25~m, and one for 50~m. 

Hence, for the same scanning rate, different models can be obtained depending on the range between the scanner and the scanned target. Furthermore, as the distance increases, the range noise also increases, even for targets exhibiting the same intensity when observed at different ranges. On the other hand, the models cannot be directly compared between the scanners due to the different scaling functions used and the differing definitions of the intensity spectrum.

In the case of the Leica~ScanStation~P50, as shown in \autoref{fig:Intensity-based range variance models from scaled intensity values_laboratory data set_Leica P50}, the model parameters $a$, $b$, and $c$ are estimated starting from 20~m (\autoref{tab:intensity_based_range_variance_parameters_from_P50}). This choice is made to avoid close-range fluctuations in the intensity values that are known from previous investigations. In contrast, the scaled intensity values provided by the Z+F~Imager~5016A allow a reliable estimation of the model parameters even at a distance of 10~m (\autoref{tab:intensity_based_range_variance_parameters_from_ZF}), indicating a more stable scaling behavior at short ranges.

\begin{figure*}
    \centering
    {%
    \captionsetup[subfigure]{justification=centering}
    \begin{subfigure}[t]{0.47\textwidth}
        \centering
        \includegraphics[width=\linewidth]{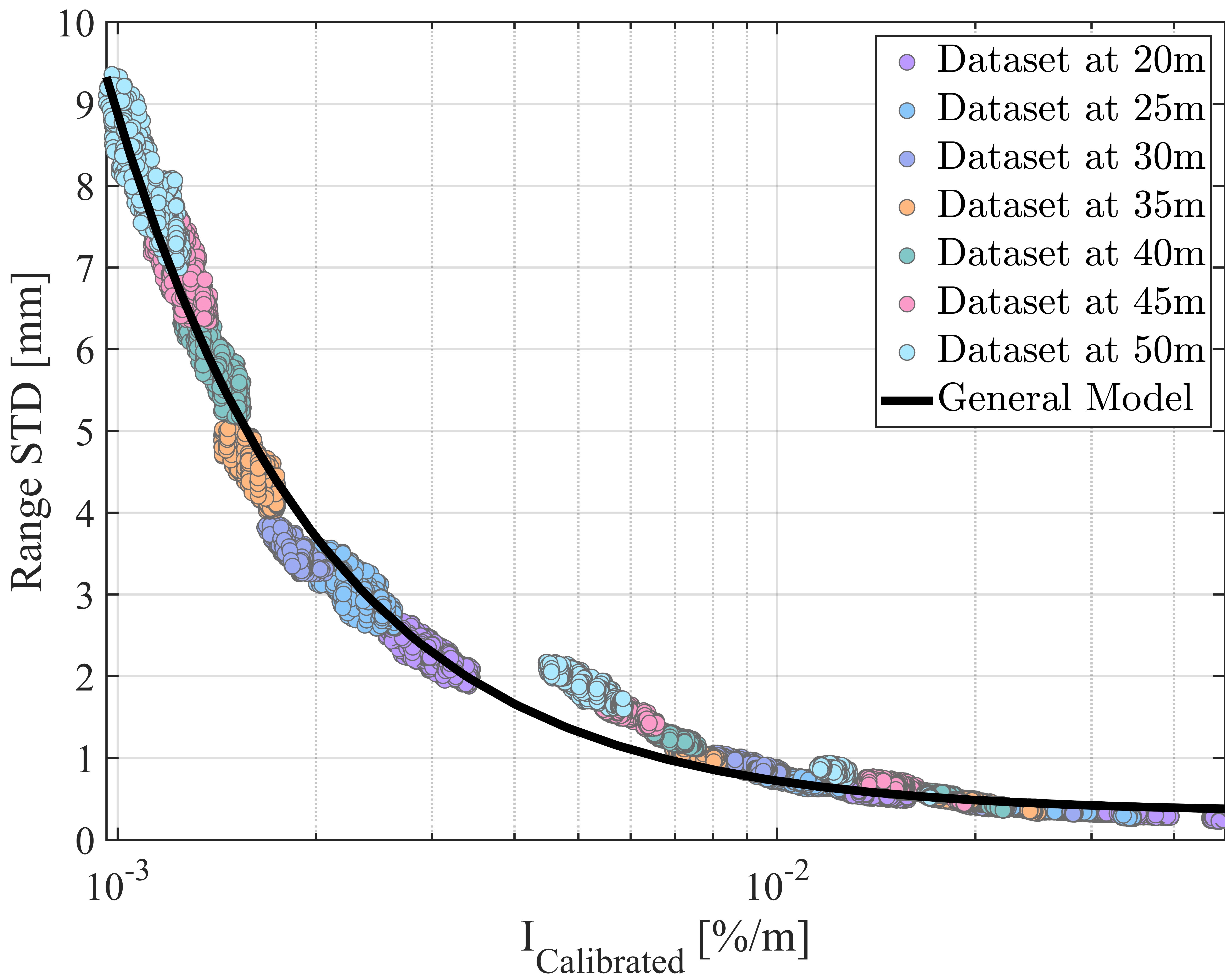}
        \caption{}
        \label{fig:General_Intensity-based range variance models from scaled intensity values_laboratory dataset_Leica P50}
    \end{subfigure}
    \hfill
    \begin{subfigure}[t]{0.47\textwidth}
        \centering
        \includegraphics[width=\linewidth]{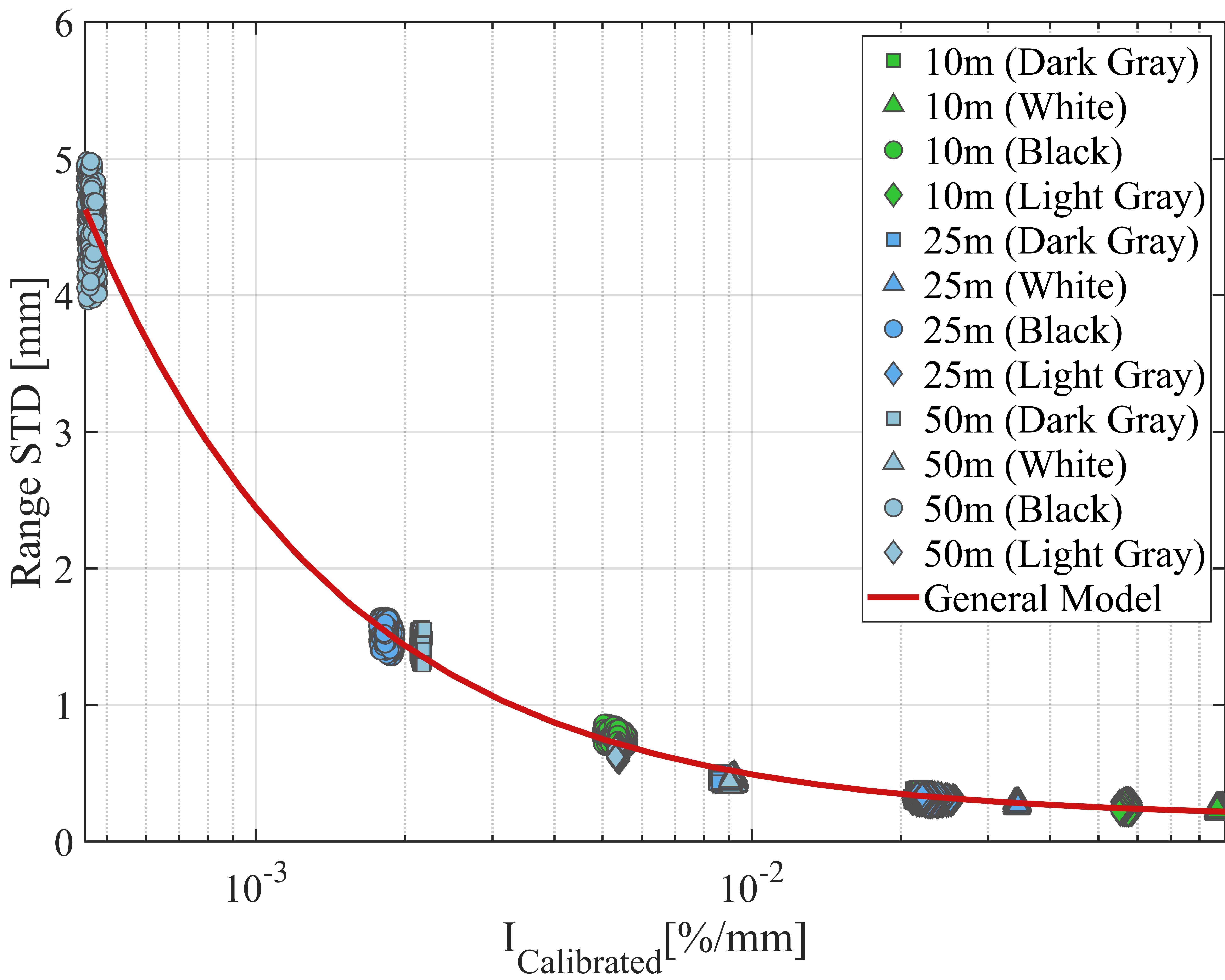}
        \caption{}
        \label{fig:General_Intensity-based range variance models from scaled intensity values_laboratory dataset_Z+FImager5016A}
    \end{subfigure}
    }
    \caption{General intensity-based range variance models derived from calibrated scaled intensity data acquired with the Leica~ScanStation~P50 and the Z+F~Imager~5016A. (a) Model estimation based on Leica~ScanStation~P50 data sets acquired at distances between 20~m and 50~m. (b) Model estimation based on Z+F~Imager~5016A data sets acquired at 10~m, 25~m, and 50~m.}

    \label{fig:General_model_comparison_scaled intensities_p50_ZF}
\end{figure*}

\begin{center}
\captionof{table}{Estimated parameters for the intensity-based range variance models derived from data sets acquired at different distances while scanning the Spectralon board using the  Leica~ScanStation~P50.}
\label{tab:intensity_based_range_variance_parameters_from_P50}
\begin{tabular}{llll}
\starttabularbody
Distance [m] & $a$ [$\frac{\text{mm}}{\text{\%}}$] & $b$ [--] & $c$ [mm] \\ \midrule
20 & 0.10 & -1.04 & 0.13 \\
25 & 0.18 & -0.99 & 0.08 \\
30 & 0.31 & -0.84 & -0.03 \\
35 & 0.27 & -0.97 & 0.06 \\
40 & 0.37 & -0.96 & 0.00 \\
45 & 0.47 & -0.94 & -0.06 \\
50 & 0.56 & -0.93 & -0.09 \\
\end{tabular}
\end{center}

\begin{center}
\captionof{table}{Intensity-based range variance model parameters derived from Z+F~Imager~5016A observations of the Spectralon board at 10~m, 25~m, and 50~m.}
\label{tab:intensity_based_range_variance_parameters_from_ZF}
\begin{tabular}{llll}
\starttabularbody
Distance [m] & $a$ [$\frac{\text{mm}}{\text{\%}}$] & $b$ [--] & $c$ [mm] \\ \midrule
10 & 0.03 & -0.98 & 0.20 \\
25 & 0.05 & -1.08 & 0.22 \\
50 & 0.49 & -0.73 & -0.09 \\
\end{tabular}
\end{center}
\subsubsection{Model estimation based on calibrated intensity}

In order to overcome the effect of intensity scaling, we estimate the generalized intensity-based range variance model described in Section~\ref{General model for scaled intensity}. This approach requires a reference range, denoted as $r_{\text{ref}}$, which is set as the average range for each individual data set (i.e., 20~m, 25~m, 30~m, 35~m, 40~m, 45~m, 50~m, for  Leica~ScanStation~P50; 10~m, 25~m, for Z+F~Imager~5016A). Accordingly, a calibrated mean intensity is estimated for each vertical tick. These calibrated intensity values, together with the corresponding range uncertainties, are then used to estimate the general model parameters $a$, $b$, and $c$ 

Instead of requiring a separate intensity-based range variance model for each specific distance at the same scanning rate, the proposed general model can be applied across all ranges, as shown in \autoref{fig:General_model_comparison_scaled intensities_p50_ZF}. Therefore, given the scaled intensity, the measuring distance, and the corresponding reference range, the calibrated intensity can be computed to further derive the range variance using the general model. 

While the model fits very well in \autoref{fig:General_Intensity-based range variance models from scaled intensity values_laboratory dataset_Z+FImager5016A}, it can also be observed in \autoref{fig:General_Intensity-based range variance models from scaled intensity values_laboratory dataset_Leica P50} that the model does not fit well in the middle section. This is because the calibration function in Eq. (\ref{eq:calibrated_intensity}) is based only on the range values, while the scaling function used by the Leica~ScanStation~P50 most probably takes into account not only the range but also other parameters, such as the incidence angles and additional factors defined by the manufacturer, which are not publicly known. 

To overcome this limitation, the calibration function of Eq. (\ref{eq:calibrated_intensity}) could be extended by including additional parameters, such as the incidence angles according to Eq. (\ref{eq:radar_range_equation}). However, on the one hand, this extension does not improve the results consistently, based on our investigations. On the other hand, our goal is to keep the proposed function as simple as possible, considering a wide applicability among different laser scanners having in mind that the range values are the most dominant factor in the intensity calibration process. For incidence angles above approximately 40 degrees, extended incidence angle corrections could become more relevant. Anyhow, since we currently cannot achieve high-quality measurements in those cases, this scenario is at the moment outside of the scope of our investigations. 

\subsubsection{Evaluation of the model derived from scaled intensities}

In the previous subsection, we introduced two approaches for estimating the intensity-based range variance model from scaled intensity values: either by deriving a separate model for each distance at the same scanning rate or by calibrating the intensity values to obtain one general model applicable to all distances. In this subsection, we focus on evaluating models estimated for individual distances. Two evaluation procedures are introduced. The first evaluates the intensity-based range variance models estimated at 20~m and 30~m from the laboratory data sets observed with the Leica~ScanStation~P50 with point spacing of 3.1~mm at 10~m using the Brucher Water Dam data sets acquired at the same distances under identical scanning configurations. The second evaluation procedure involves estimating two models from the Brucher Water Dam data sets and subsequently comparing them with the corresponding laboratory-derived models at 20~m and 30~m.

The first approach focuses on evaluating the laboratory-derived models using the Brucher Water Dam data sets. The data were acquired at 20~m and 30~m using the Leica~ScanStation~P50. For each of these data sets, we apply the preprocessing workflow described in Section~\ref{Workflow for data preprocessing}. Accordingly, for each vertical tick, we compute the mean intensities \(\bar{I}_{Env}^{20m}\), \(\bar{I}_{Env}^{30m}\) and the corresponding range standard deviations \(\sigma_{r,Env}^{20m}\), \(\sigma_{r,Env}^{30m}\). Using the same intensity values as input to the intensity-based range variance models derived from the laboratory data sets at 20~m and 30~m, we estimate the corresponding predicted range standard deviations \(\hat{\sigma}_{r,Lab}^{20m}\) and \(\hat{\sigma}_{r,Lab}^{30m}\). Finally, we compute the residuals as the difference between the predicted range standard deviations \(\hat{\sigma}_{r,Lab}^{20m}\), \(\hat{\sigma}_{r,Lab}^{30m}\) and those obtained from the preprocessing step, \(\sigma_{r,Env}^{20m}\), \(\sigma_{r,Env}^{30m}\). Based on these residuals, we determine the RMSE and the maximum absolute residuals, as shown in \autoref{tab:validation_results_scaled_intensity_crossvalidation}.

\begin{table}
\caption{RMSE and maximum absolute residuals between the range standard deviations predicted by the intensity-based range variance models derived from the Laboratory datasets (at 20~m and 30~m), and the corresponding standard deviations estimated from the Brucher Water Dam datasets (at 20~m and 30~m) after data preprocessing.}
\centering
\begin{tabular}{lll}
\starttabularbody
Evaluation data sets          & RMSE [mm] & Max. Residuals [mm] \\ \midrule
20~m              & 0.07      & 0.26                \\
30~m              & 0.10      & 0.29                \\
\end{tabular}
\label{tab:validation_results_scaled_intensity_crossvalidation}
\end{table}

In addition to cross-validation, the second validation approach is conducted by estimating models directly from the observed data. We apply the model fitting procedure outlined in Section~\ref{Workflow for model fitting} to the preprocessed Brucher Water Dam data sets acquired at 20~m and 30~m. As a result, two sets of model parameters are estimated for these two distances, as presented in \autoref{tab:intensity_based_range_variance_parameters_from_natural_targets_P50}. Each observation-derived model is then compared to its corresponding laboratory-derived model at 20~m and 30~m. This comparison is based on evaluating the residuals between the range standard deviations predicted by the two models using the same mean scaled intensity values. Finally, we compute the RMSE and the maximum absolute residuals for both distances, as shown in \autoref{tab:validation_results_scaled_intensity_model_comparison_scaled_intensity}.

\begin{center}
\includegraphics[width=\linewidth]{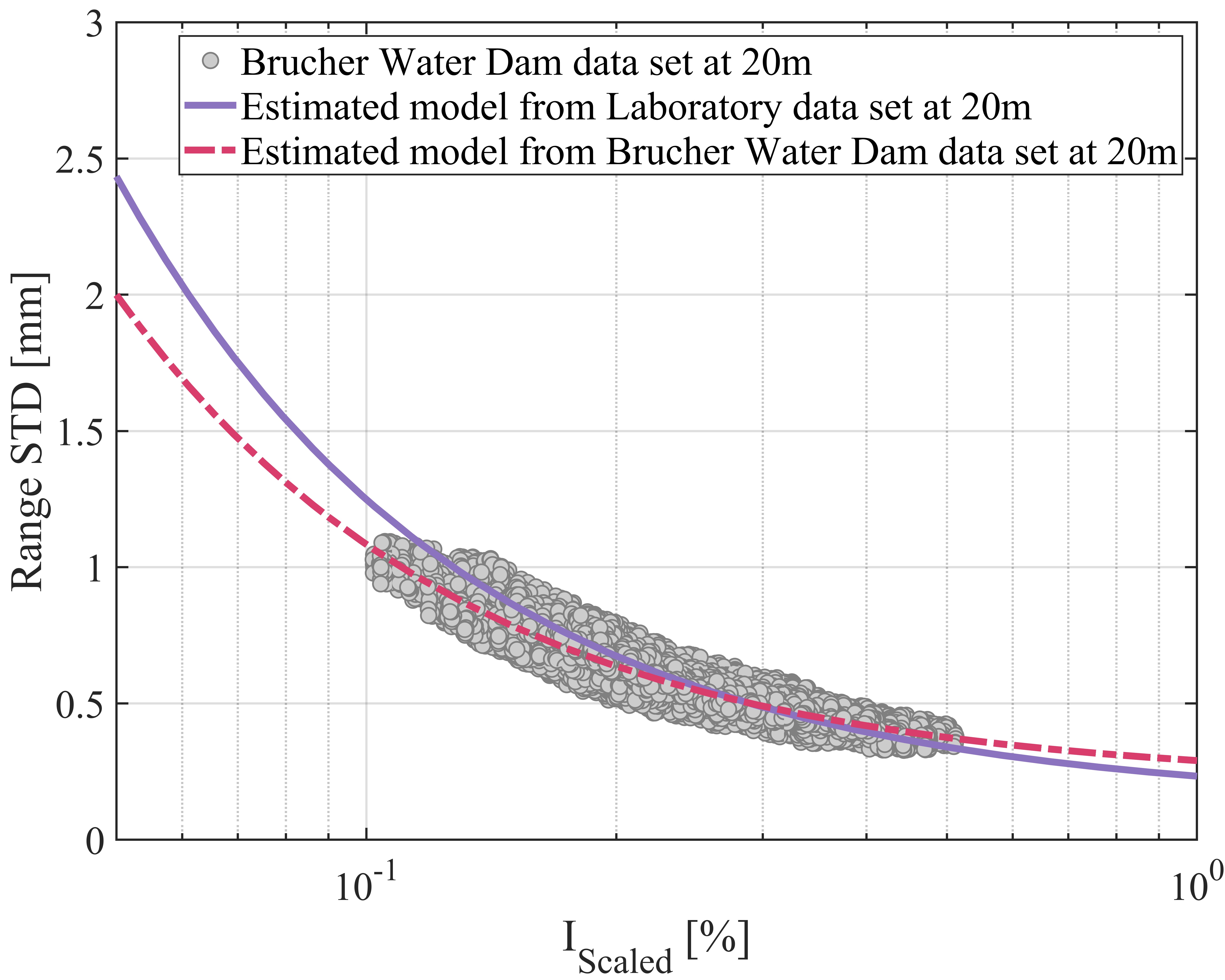}
\captionof{figure}{Comparison of intensity-based range variance models derived from the laboratory data set at 20~m (purple solid curve) and the Brucher Water Dam data set at 20~m (red dashed curve)  observed with the Leica~ScanStation~P50 with point spacing of 3.1~mm at 10~m.}
\label{fig:intensity_based_range_variance_model_comparison_P50_20m}
\end{center}

\vspace{2em}

\begin{center}
\captionof{table}{Model parameters of the intensity-based range variance models derived from the scaled intensity values of the Brucher Water Dam data sets at 20~m and 30~m.}
\label{tab:intensity_based_range_variance_parameters_from_natural_targets_P50}
\begin{tabular}{llll}
\starttabularbody
Evaluation data sets & $a$ [$\frac{\text{mm}}{\text{\%}}$] & $b$ [--] & $c$ [mm] \\ \midrule
Water dam at 20 & 0.06 & -1.13 & 0.24 \\
Water dam at 30 & 0.47 & -0.64 & -0.17 \\
\end{tabular}
\end{center}

\begin{center}
\includegraphics[width=\linewidth]{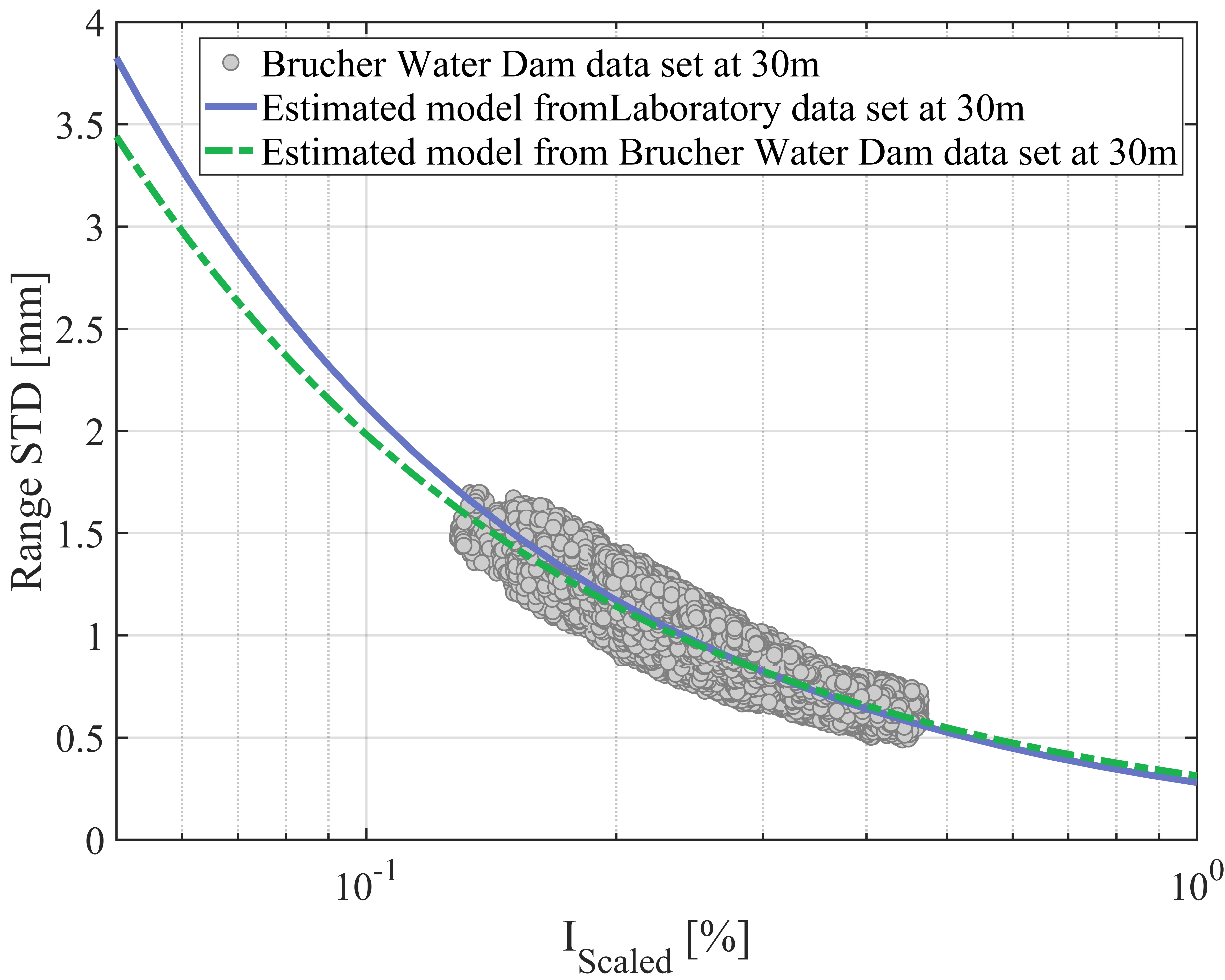}
\captionof{figure}{Comparison of intensity-based range variance models derived from the laboratory data set at 30~m (blue solid curve) and the Brucher Water Dam data set at 30~m (green dashed curve) observed with the Leica~ScanStation~P50 with a point spacing of 3.1~mm at 10~m.}
\label{fig:intensity_based_range_variance_model_comparison_P50_30m}
\end{center}

\begin{center}
\captionof{table}{RMSE and maximum absolute residuals between the range standard deviations predicted by the intensity-based range variance model estimated from the Laboratory datasets at 20~m and 30~m, and the corresponding standard deviations derived from the model estimated from the Brucher Water Dam datasets at the same distances.}
\label{tab:validation_results_scaled_intensity_model_comparison_scaled_intensity}
\begin{tabular}{lll}
\starttabularbody
Evaluation  data sets          & RMSE [mm] & Max. Residuals [mm] \\ \midrule
20~m              & 0.04      & 0.16                \\
30~m              & 0.02      & 0.09                \\
\end{tabular}
\end{center}

By comparing the RMSE values and the maximum absolute residuals reported in Tables \ref{tab:validation_results_scaled_intensity_crossvalidation} and \ref{tab:validation_results_scaled_intensity_model_comparison_scaled_intensity}, it becomes evident that these values are notably reduced when using models derived directly from the scanned data sets. This indicates that utilizing models estimated from the actual measurement environment improves the estimation of range uncertainties, but only within the intensity spectrum covered by the object’s intensity values, compared to relying solely on models developed from laboratory observations. Furthermore, as shown in \autoref{tab:validation_results_scaled_intensity_model_comparison_scaled_intensity}, the RMSE at 20~m is higher than that at 30~m. This difference can be attributed to the increased range noise at closer distances, where the concave geometry and surface complexity of the water dam have a more pronounced influence, as illustrated in \cref{fig:intensity_based_range_variance_model_comparison_P50_20m,fig:intensity_based_range_variance_model_comparison_P50_30m}.

\subsubsection{Evaluation of the model derived from calibrated intensities values}

After evaluating the intensity-based range variance models derived for each individual distance, we evaluate the general model estimated for the Leica~ScanStation~P50 using the same scanning resolution. Two evaluation approaches are used to assess this model. The first approach involves comparing the model shown in \autoref{fig:General_Intensity-based range variance models from scaled intensity values_laboratory dataset_Leica P50} with the separate models in \autoref{fig:Intensity-based range variance models from scaled intensity values_laboratory data set_Leica P50} using the laboratory data sets acquired at distances from 20~m to 50~m. On the other hand, the second approach is based on the Brucher Water Dam data sets acquired at 20~m and 30~m using the same scanner and the same point spacing as in the first approach.

The first evaluation approach examines how well the general model represents the range uncertainties compared to the individual models derived for each distance. The range standard deviations predicted by the general model are compared with those obtained from the individual intensity-based range variance models for each distance from the laboratory datasets. For this comparison, the preprocessed intensity values are calibrated for each Spectralon color at each distance using Eq. (\ref{eq:calibrated_intensity}). The RMSE and the maximum residual values are then calculated as presented in \autoref{tab:validation_results_general_vs_lab_scaled_intensity_using_spectralon_all_ranges}.

\begin{center}
\captionof{table}{General model validation by comparing RMSE and maximum absolute residual between the range standard deviations predicted by the Laboratory-based models (evaluated at distances from 20~m to 50~m in 5~m intervals) and those derived from the general model using calibrated intensity values.}
\label{tab:validation_results_general_vs_lab_scaled_intensity_using_spectralon_all_ranges}
\begin{tabular}{lll}
\starttabularbody
Evaluation data sets          & RMSE [mm]           & Max. Residuals [mm] \\ \midrule
20                            & 0.09                & 0.13 \\
25                            & 0.10                & 0.29 \\
30                            & 0.29                & 0.80 \\
35                            & 0.20                & 0.56 \\
40                            & 0.16                & 0.32 \\
45                            & 0.28                & 0.50 \\
50                            & 0.39                & 0.67 \\
\end{tabular}
\end{center}
\vspace{2em}

The second evaluation approach further assesses the general model’s applicability under in situ conditions. First, the range standard deviations for the Brucher Water Dam data sets at 20~m and 30~m are estimated by applying the individual intensity-based range variance models generated from the laboratory data sets at the same distances. Afterwards, the general model is applied to the same data sets after calibrating the intensity values corresponding to each distance. Hence, the residuals between the standard deviations estimated from the two approaches are computed for each distance. The RMSE and the maximum absolute residuals are then calculated, as shown in \autoref{tab:validation_results_general_vs_lab_scaled_intensity_using_waterdam}.

\begin{center}
\captionof{table}{RMSE and maximum absolute residuals between the range standard deviations predicted by the laboratory data sets at 20~m and 30~m and general models, using the preprocessed Brucher Water Dam data sets at 20~m and 30~m.}
\label{tab:validation_results_general_vs_lab_scaled_intensity_using_waterdam}
\begin{tabular}{lll}
\starttabularbody
Validation  data sets          & RMSE [mm] & Max. Residuals [mm] \\ \midrule
20~m                          & 0.11      & 0.35                \\
30~m                          & 0.13      & 0.34                \\
\end{tabular}
\end{center}
\vspace{2em}

The results in Tables \ref{tab:validation_results_general_vs_lab_scaled_intensity_using_spectralon_all_ranges} and \ref{tab:validation_results_general_vs_lab_scaled_intensity_using_waterdam} indicate that RMSE values remain below one millimeter, confirming that the general model is a reliable alternative for estimating range uncertainties in scanners providing scaled intensity values.

\subsubsection{Summarizing discussion}

The models estimated from the scaled intensity data sets show that the range noise increases with distance and varies depending on the scaling functions applied by each scanner. The general model estimated from the calibrated intensity values eliminates the need for a separate model for each distance while keeping the same behavior as the models derived from the laboratory data sets. When evaluated using the Brucher Water Dam data sets, the laboratory-derived models fit well within the observed intensity range, confirming their applicability and transferability to real-world measurement conditions. The models estimated from the real-world data also work well within the range of the intensity spectrum used for their derivation; otherwise, extrapolation effects occur. The evaluation of the general model using both the laboratory and Brucher Water Dam data sets shows that it performs consistently across all ranges, confirming its reliability and stability.

\section{Conclusion}
\label{conclusion}
This article introduces a workflow for estimating intensity-based range variances of TLS for scanners providing either raw intensity values (e.g., Z+F~Imager~5016A) or scaled intensities (e.g.,  Leica~ScanStation~P50), using a 2D scanning mode. The method was first tested in a controlled laboratory with a Spectralon board of four reflectance levels to derive range variance models at varying distances.

For raw intensity data sets, cross-validation against real-world data sets confirms the Spectralon-based models’ applicability. Results showed that smooth and uniform surfaces (e.g., Pinakothek Wall) yield lower range variances, while high curvature or roughness and non-homogeneous colors (e.g., metallic sculpture, water dam) can increase the uncertainty, highlighting the influence of surface geometry and material properties on range uncertainty estimation. 

For scaled intensity data sets, the scaling function applied during data export can alter the model behavior, requiring separate models for different measuring distances. To address this, a general model was introduced by calibrating intensities based on range at each vertical angle tick, enabling a single model valid over the full range.

Evaluation using the Brucher Water Dam data sets at 20~m and 30~m showed sub-millimeter RMSEs and maximum absolute residuals for both the raw and scaled intensity-based models, confirming their feasibility and reliability. Model parameters ($a$, $b$, $c$) vary across data sets, reflecting their dependence on specific surface properties. Differences between laboratory-based and in-situ models (below 1~mm) were negligible within the tested intensity spectrum. However, models derived directly from in-situ objects may remain surface-dependent and are not universally transferable across all intensity ranges due to extrapolation effects.


\begin{funding}
This research was funded by the German Research Foundation (DFG) under grant number 490989047, DFG FOR 5455 "TLS-Defo".
\end{funding}

\bibliographystyle{plainnat}
\bibliography{Bibliography}
\end{document}